\documentclass[journal]{IEEEtran}
\usepackage{amssymb}
\usepackage{caption2}
\usepackage{amsmath}
\usepackage{multirow}
\usepackage{amsthm}
\usepackage[ruled,vlined,commentsnumbered]{algorithm2e}
\newtheorem{theorem}{Theorem}[section]
\newtheorem{definition}[theorem]{Definition}
\newtheorem{lemma}[theorem]{Lemma}
\newtheorem{corollary}[theorem]{Corollary}

\begin{document}

\title{Multimedia IPP Codes with Efficient Tracing}
\author{Jing Jiang, ~Minquan Cheng, ~Ying Miao and Dianhua Wu
\thanks{The research of Cheng was supported in part by NSFC (No.~11301098),
Guangxi Natural Science Foundation (No.~2013GXNSFCA019001), and the Scientific Research Foundation for the Returned Overseas Chinese Scholars, State Education Ministry.
The research of Miao was supported by JSPS Grant-in-Aid for Scientific Research (C) under Grant No.~24540111.
The research of Wu was supported in part by  NSFC (No.~11271089), Guangxi Natural
Science Foundation (No.~2012GXNSFAA053001; No.~2014GXNSFDA1180001), Program on the High Level
Innovation Team and Outstanding Scholars in Universities of Guangxi Province,
and Guangxi ``Ba Gui" Team for Research and Innovation.
}

\thanks{J. Jiang and Y. Miao are with the Department of Social Systems and Management,
Graduate School of Systems and Information Engineering, University of Tsukuba, Tsukuba 305-8573, Japan.
E-mails: jjiang2008@hotmail.com, miao@sk.tsukuba.ac.jp.}

\thanks{M. Cheng and D. Wu are with the School of Mathematics and Statistics, Guangxi Normal University, Guilin 541004, P. R. China.
E-mails: chengqinshi@hotmail.com,  dhwu@gxnu.edu.cn.}

\thanks{Copyright (c) 2014 IEEE. Personal use of this material is permitted. However, permission to use this material for any other purposes must be obtained from the IEEE by sending a request to pubs-permissions@ieee.org.}
}
\maketitle

\begin{abstract}
Binary multimedia identifiable parent property codes (binary $t$-MIPPCs) are used
in multimedia fingerprinting schemes where the identification of users taking part in the averaging collusion attack to
illegally redistribute content is required.
In this paper, we first introduce a binary strong
multimedia identifiable parent property code (binary $t$-SMIPPC) whose
tracing algorithm is more efficient than that of a binary $t$-MIPPC.
Then a composition construction for binary $t$-SMIPPCs from $q$-ary $t$-SMIPPCs is provided.
Several infinite series of optimal $q$-ary $t$-SMIPPCs  of length $2$ with $t = 2, 3$ are derived from
the relationships among $t$-SMIPPCs and other fingerprinting codes,
such as $\overline{t}$-separable codes and $t$-MIPPCs.
Finally, combinatorial properties of $q$-ary $2$-SMIPPCs of length $3$ are investigated,
and optimal $q$-ary $2$-SMIPPCs  of length $3$ with $q \equiv 0, 1, 2, 5 \pmod 6$ are constructed.
\end{abstract}


\section{Introduction}
\label{intro}
%
%
%
%

\IEEEPARstart{C}odes with the identifiable parent property ($t$-IPPCs) were
first introduced by Hollmann et al. \cite{HLLT}, motivated
by the purpose of protecting copyrighted digital contents,
and investigated in detail in \cite{BCEKZ, BK, Bl, SSW, TM}.
Recently, Cheng et al. \cite{CFJLM} introduced a multimedia identifiable parent property code ($t$-MIPPC)
to resist the averaging collusion attack on multimedia contents.
They showed that binary $t$-MIPPCs can be used in multimedia fingerprinting to identify,
as $t$-IPPCs do in the generic digital scenario \cite{BBK, HLLT},
at least one such malicious authorized user when the size of the coalition is at most $t$ with
computational complexity $O(nM^{t})$, where $n$ is the code length and $M$ is the number of codewords,
thereby bringing enough pressure to bear on authorized users to give up their attempts at collusion.
Obviously, the tracing algorithm based on such a binary $t$-MIPPC is not efficient for practical use. In this paper, we
introduce a new notion of a strong  multimedia identifiable parent property code ($t$-SMIPPC) to resist
the averaging attack on multimedia contents in a fingerprinting system. We show that binary $t$-SMIPPCs can be used in tracing algorithms to identify
at least one colluder when the size of the coalition is at most $t$ with
computational complexity $O(nM)$, which is clearly more efficient than those based on binary $t$-MIPPCs.

This paper is organized as follows.
In Section \ref{pre}, we briefly review the concepts of fingerprinting, collusion and detection.
In Section \ref{SMIPPC}, we introduce the notion of a strong multimedia identifiable parent property code,
and discuss the tracing algorithm based on this new code.
In Section \ref{optimalSMIPPC}, several infinite series of optimal $t$-{\rm SMIPPC}s of length $2$
with $t = 2, 3$ are derived.
In Section \ref{construction}, optimal $q$-ary $2$-SMIPPCs  of length $3$ with $q \equiv 0, 1, 2, 5 \pmod 6$ are constructed.
Finally, conclusions will be given in Section VI.

\section{Preliminaries } %
\label{pre}                                                                  %

In this section, we give a very brief review on some basic terminologies.
The interested reader is referred to \cite{CM,LTWWZ} for more detailed information.

In collusion-resistant fingerprinting, we want to design fingerprints which can be used to
trace and identify colluders after collusion attacks, together with robust embedding of
fingerprints into multimedia host signals. Spread-spectrum additive embedding is one of
the widely employed robust embedding techniques.
Let ${\bf x}$ be the host multimedia signal, $\{ {\bf u}_i \ | \ 1 \leq i \leq n\}$ be an
orthonormal basis of noise-like signals, and $\{{\bf w}_j = ({\bf w}_j(1), {\bf w}_j(2),
\ldots, {\bf w}_j(n)) = \sum_{i=1}^{n}b_{ij}{\bf u}_i\ | \ 1 \leq j \leq M\}$, $b_{ij} \in \{0,1\}$,
be a family of scaled watermarks to achieve the imperceptibility as well as to control the energy of the embedded watermark.
The watermarked version of the content ${\bf y}_j = {\bf x} + {\bf w}_j$, $1 \leq j \leq M$, is then assigned to
the authorized user $U_j$ who has purchased the rights to access ${\bf x}$.
The fingerprint ${\bf w}_j$  assigned to $U_j$ can be represented uniquely by a vector (called codeword)
${\bf b}_j = (b_{1j}, b_{2j}, \ldots, b_{nj})^{T} \in \{0,1\}^{n}$ because of the linear independence
of the basis $\{ {\bf u}_i \ | \ 1 \leq i \leq n\}$.

When $t$ authorized users, say $U_{j_1}, U_{j_2}, \ldots, U_{j_t}$,
who have the same host content  but different fingerprints come together,
we assume that they have no way of manipulating the individual orthonormal signals,
that is, the underlying codeword needs to be taken and proceeded as a single entity,
but they can carry on a linear collusion attack to generate a pirate copy from their $t$ fingerprinted contents,
so that the venture traced by the pirate copy can be attenuated. In additive embedding,
this is done by linearly combining the $t$ fingerprinted contents
$\sum_{l=1}^{t}\lambda_{j_l}{\bf y}_{j_l}$, where the weights $\{{\lambda}_{j_l} \ | \ 1 \leq l \leq t\}$
satisfy the condition $\sum_{l=1}^{t}\lambda_{j_l} = 1$ to maintain the average intensity of the original multimedia signal.
In this case, the energy of each of the watermarks ${\bf w}_{j_l}$ is reduced by a factor of $\lambda_{j_l}^{2}$,
therefore, the trace of $U_{j_l}$'s fingerprint becomes weaker and thus $U_{j_l}$ is less likely to be caught by the detector.
Since normally no colluder is willing to take more of a risk than any other colluder,
averaging attack in which ${\lambda}_{j_l} = 1/t$, $1 \leq l \leq t$, is the most fair choice
for each colluder to avoid detection, as claimed in \cite{LTWWZ,TWWL}.
This attack also makes the pirate copy have better perceptional quality.
Based on the discussions above, the observed content ${\bf y}$ after averaging attack  is
\[ {\bf y} = \frac{1}{t}\sum\limits_{l=1}^{t}{\bf y}_{j_l} = \frac{1}{t}\sum\limits_{l=1}^{t}{\bf w}_{j_l} + {\bf x}=
 \sum\limits_{l=1}^{t}\sum\limits_{i=1}^{n}\frac{b_{ij_l}}{t}{\bf u}_{i} + {\bf x}.\]

In colluder detection phase,
we compute the correlation vector
${\bf T} = ( {\bf T}(1), {\bf T}(2), \ldots, {\bf T}(n))$,
where ${\bf T}(i) = \langle {\bf y}-{\bf x}, {\bf u}_{i}\rangle$, $1 \leq i \leq n$,
and $\langle {\bf y}-{\bf x}, {\bf u}_{i}\rangle$ is the inner product of ${\bf y}-{\bf x}$ and  ${\bf u}_i$.
We would like to strategically design an anti-collusion code  to accurately identify the contributing fingerprints
involved in the averaging attack from this detection statistics ${\bf T}$.

\section{Strong Multimedia Identifiable Parent Property Codes}   %
\label{SMIPPC}

In this section, we first introduce the notion of a strong multimedia identifiable parent property code ($t$-SMIPPC),
and then discuss the tracing algorithm based on binary SMIPPCs. A composition construction for  binary $t$-SMIPPCs
from $q$-ary $t$-SMIPPCs is also presented.

Let $n, M$ and $q$ be positive integers, and $Q$ an alphabet with $|Q|=q$.
A set ${\cal C} = \{{\bf c}_1,{\bf c}_2,\ldots, {\bf c}_M\} \subseteq Q^n$ is called an $(n,M,q)$ code
and each ${\bf c}_i$ is called a codeword.
Without loss of generality, we may assume $Q=\{0,1,\ldots,q-1\}$.
When $Q=\{0,1\}$, we also use the word ``binary".
Given an $(n,M,q)$ code, its incidence matrix is the $n \times M$ matrix on $Q=\{0,1,\ldots,q-1\}$
in which the columns are the $M$ codewords in ${\cal C}$.
Often, we make no difference between an $(n,M,q)$ code and its incidence matrix unless otherwise stated.

For any $(n,M,q)$ code $\cal{C}$ on $Q$, we define
the following shortened code ${\cal A}_{i}^{j}$ for $i \in Q$ and $1 \leq j \leq n$:
\begin{eqnarray*}
 {\cal A}_{i}^{j} = \{({\bf c}(1), \ldots, {\bf c}(j-1), {\bf c}(j+1), \ldots, {\bf c}(n))^{T} \ | \\
({\bf c}(1), \ldots, {\bf c}(n))^{T} \in {\cal C}, {\bf c}(j) = i\}.
\end{eqnarray*}

For any code ${\cal C} \subseteq Q^n$, we define the set of $i$th coordinates of ${\cal C}$ as
\[ {\cal C}(i) =\{{\bf c}(i) \in Q \ | \ {\bf c}=({\bf c}(1), {\bf c}(2), \ldots, {\bf c}(n))^{T} \in {\cal C}\}\]
for any $1 \le i \le n$.
For any subset of codewords ${\cal C}^{'} \subseteq {\cal C}$, we define the descendant code of ${\cal C}^{'}$ as
\begin{eqnarray*}
 {\sf desc}({\cal C}^{'}) = \{({\bf x}(1), {\bf x}(2), \ldots, {\bf x}(n) )^{T}  \in Q^n \ |   \\  {\bf x}(i) \in {\cal C}^{'}(i), 1 \le i \le n\},\end{eqnarray*}
that is,
\[ {\sf desc}({\cal C}^{'})={\cal C}^{'}(1) \times {\cal C}^{'}(2) \times \cdots \times  {\cal C}^{'}(n).\]
The set ${\sf desc}({\cal C}^{'})$ consists of the $n$-tuples that could be
produced by a coalition holding the codewords in ${\cal C}^{'}$.

Using the notions of descendant codes and sets of $i$th coordinates of codes, Cheng et al. \cite{CFJLM} defined multimedia
identifiable parent property codes (MIPPCs) and discussed the tracing algorithm based on binary MIPPCs.

\begin{definition}
\label{MIPP}
Let $\cal C$ be an $(n,M,q)$ code, and for any $R \subseteq {\cal C}(1) \times {\cal C}(2) \times \cdots \times {\cal C}(n)$,
define the set of parent sets of $R$ as
\[ {\cal P}_{t}(R) = \{{\cal C}^{'} \subseteq {\cal C} \ | \ |{\cal C}^{'} | \leq t, {\sf desc}({\cal C}^{'}) = R\}. \]
We say that $\cal C$ is a code with the identifiable parent property (IPP) for multimedia fingerprinting, or a multimedia IPP code, denoted $t$-MIPPC$(n,M,q)$,
if $\bigcap_{{\cal C}^{'} \in {\cal P}_{t}(R)}{\cal C}^{'} \neq \emptyset$
is satisfied for any $R \subseteq {\cal C}(1) \times {\cal C}(2) \times \cdots \times {\cal C}(n)$ with ${\cal P}_{t}(R) \neq \emptyset$.
\end{definition}

The notion of a binary MIPPC was introduced in \cite{CFJLM}
for protecting multimedia contents, which, with code modulation, can be used to
construct families of fingerprints with the ability to survive collusion and trace colluders.
In fact, in the multimedia scenario,
for any set of colluders holding codewords ${\cal C}_0 \subseteq {\cal C}$ and for any index $1 \leq i \leq n$,
their detection statistics ${\bf T}(i)$ mentioned in Section \ref{pre} possesses the whole information
on ${\cal C}_0(i)$, namely, we have ${\bf T}(i)=1$ if and only if ${\cal C}_0(i)=\{1\}$,
${\bf T}(i)=0$ if and only if ${\cal C}_0(i)=\{0\}$, and
$0 < {\bf T}(i) < 1$ if and only if ${\cal C}_0(i)=\{0,1\}$.
Therefore,
we can capture a set $R = {\cal C}_0(1) \times \cdots \times {\cal C}_0(n)
 \subseteq {\cal C}(1) \times \cdots \times {\cal C}(n)$ in the multimedia
scenario from the detection statistics ${\bf T}$. The property an MIPPC holds makes it
easy to identify ${\cal C}_0$, and thus the set of colluders holding ${\cal C}_0$
who have produced $R$.

\begin{theorem}$(${\rm \cite{CFJLM}}$)$
\label{ELACCAlg}
Under the assumption that the number of colluders in the averaging attack is at most $t$,
any $t$-MIPPC$(n, M, 2)$ can be used to identify at least one colluder  with computational complexity $O(nM^{t})$
by applying Algorithm {\tt MIPPCTraceAlg}$({R})$ described in {\rm \cite{CFJLM}}.
\end{theorem}
\begin{algorithm}[h]
\caption{ {\tt MIPPCTraceAlg}$({R})$}
\label{al93}
Given $R$\;
Find\ ${\cal P}_{t}(R) = \{{\cal C}^{'} \subseteq {\cal C} \ | \ |{\cal C}^{'} | \leq t, {\sf desc}({\cal C}^{'}) = R\}$\;

Compute ${\cal C}_0 =\bigcap_{{\cal C}^{'} \in {\cal P}_{t}(R)}{\cal C}^{'}$\;\vskip 0.2cm
\eIf{$|{\cal C}_0| \leq t$ }
{{\bf output} ${\cal C}_0$ as the set of colluders\;}{{\bf output} ``the set of colluders has size at least $t+1$"\;}
\end{algorithm}


As we can see from the theorem above, the computational complexity of the algorithm based on binary MIPPCs is
not efficient for practical use. Therefore, it is desirable to find some special MIPPCs with
efficient tracing ability.

\begin{definition}
\label{defSESC}
Let $\cal C$ be an $(n, M, q)$ code, and $t \geq 2$ be an integer. $\cal C$ is a strong
multimedia identifiable parent property code, or $t$-SMIPPC$(n, M, q)$, if for any
${\cal C}_0 \subseteq \cal C$, $1 \leq |{\cal C}_0| \leq t$,  we have $\bigcap_{{\cal C}^{'} \in S({\cal C}_0)}{\cal C}^{'}\neq \emptyset$,
where $S({\cal C}_0) = \{ {\cal C}^{'} \subseteq {\cal C} \ | \ {\sf desc}({\cal C}^{'}) = {\sf desc}({\cal C}_0) \}$.
\end{definition}

The following is an equivalent definition of an SMIPPC.

\begin{definition}
\label{defSESC2}
 Let $\cal C$ be an $(n, M, q)$ code,  and $t \geq 2$ be an integer. For any $R \subseteq {\cal C}(1) \times \cdots \times {\cal C}(n)$,
define the set of parent sets of $R$ as
\[ {\cal P}(R) = \{ {\cal C}^{'} \subseteq {\cal C} \ | \  {\sf desc}({\cal C}^{'}) = R\}. \]
We say $\cal C$ is a strong multimedia identifiable parent property code, or $t$-SMIPPC$(n, M, q)$, if
$\bigcap_{ {\cal C}^{'} \in {\cal P}(R)}{\cal C}^{'} \neq \emptyset$ is satisfied
for all $R \subseteq {\cal C}(1) \times \cdots \times {\cal C}(n)$ with ${\cal P}_{t}(R) \neq \emptyset$.
\end{definition}

We can derive the following relationship immediately from Definitions \ref{MIPP} and \ref{defSESC2}.

\begin{lemma}
\label{rela5}
Any $t$-SMIPPC$(n, M, q)$ is a $t$-MIPPC$(n, M$, $q)$.
\end{lemma}

The following theorem shows that a $t$-SMIPPC$(n,M,2)$ can be used to identify at least one colluder
in the averaging attack with computational complexity $O(nM)$, which is more efficient than that of a $t$-MIPPC$(n,M,2)$.
We in fact use Algorithm {\tt SSCTraceAlg$(R)$} presented in \cite{JCM}.  For convenience,
the detailed illustration is given below.

\begin{theorem}
\label{algSELACC}
Under the assumption that the number of colluders in the averaging attack is at most $t$,
any $t$-SMIPPC$(n, M, 2)$ can be used to identify at least one colluder  with computational complexity $O(nM)$
by applying Algorithm {\tt SSCTraceAlg$(R)$} described below.
\end{theorem}
\begin{IEEEproof} Let $\cal C$ be the  $t$-SMIPPC$(n, M, 2)$, and $R$ be the descendant code derived
from the detection statistics ${\bf T}$.
Then by applying the following tracing algorithm, Algorithm \ref{AlgSSC},
one can identify at least one colluder. The computational complexity is clearly  $O(nM)$.

\begin{algorithm}[h]
\caption{ {\tt SSCTraceAlg}(${R}$)}
\label{AlgSSC}
Define $J_a$, $J_o$ to be the sets of indices where $R(j) = \{1\}$, $R(j) = \{0\}$, respectively,
and  ${\bf J_a} = ({\bf J_a}(1), \ldots, {\bf J_a}(|J_a|))^T$,  ${\bf J_o} = ({\bf J_o}(1), \ldots, {\bf J_o}(|J_o|))^T$
to be the vector representing $R$'s coordinates where $R(j) = \{1\}$ and $R(j) = \{0\}$, respectively\;
${\bf \Phi } = {\bf 1}$\;
$U_a = \emptyset$\;
$U_o = \emptyset$\;
$U = \emptyset$\;
\For { $k=1$  {\bf to }  $|J_a|$ }
     {   $j = {\bf J_a}(k)$\;
         define ${\bf e}_j$ to be the $j$th row of $\cal C$\;
         ${\bf \Phi } = {\bf \Phi} \cdot {\bf e}_j$\;
     }

\For { $k=1$ {\bf to} $|J_o|$}
     {   $j = {\bf J_o}(k)$\;
         ${\bf \Phi } = {\bf \Phi } \cdot \overline{{\bf e}}_j$\;
     }

\For { $k = 1$ {\bf to} $n$ }
     {   ${\bf \Phi}_a = {\bf \Phi} \cdot {\bf e}_k$\;
         ${\bf \Phi}_o = {\bf \Phi} \cdot \overline{{\bf e}}_k$\;
         \For { $i = 1$ {\bf to} $M$ }
              {   \If { ${\bf \Phi}_a(i) = 1$ }
                      {   $U_a = \{i\} \bigcup U_a$\;
                      }
              }
         \If  { $|U_a|=1$ }
              {   $U = U \bigcup U_a$\;
              }
         \For { $i = 1$ {\bf to} $M$ }
              {   \If { ${\bf \Phi}_o(i) = 1$}
                      {   $U_o = \{i\} \bigcup U_o$\;
                      }
              }

         \If { $|U_o|=1$  }
             { $U = U \bigcup U_o$\;
             }
    }

\eIf { $|U| \leq t$ }
    { {\bf output }$U$\;}
    { {\bf output} ``The set of colluders has size at least $t+1$."}
\end{algorithm}

According to  Algorithm \ref{AlgSSC}, by deleting all columns
$\{ {\bf c} \in {\cal C} \  | \    \exists  \ 1 \leq i \leq n,
R(i) = \{1\}, {\bf c}(i) = 0, \ {\rm or} \ R(i) = \{0\}, {\bf c}(i) = 1\}$, we obtain a sub-matrix ${\cal C}_{L}$ of ${\cal C}$.
Suppose that $C_0 = \{u_1, u_2, \ldots, u_r\}$, $1 \leq r \leq t$,
is the set of colluders, the codeword ${\bf c}_i$ is assigned to the colluder $u_i$,  $1 \leq i \leq r$,
and ${\cal C}_0 = \{ {\bf c}_1, {\bf c}_2, \ldots, {\bf c}_r\}$.
It is not difficult to see that ${\cal C}_0 \subseteq {\cal C}_{L}$.
According to the definition of a $t$-SMIPPC,
we have $\bigcap_{{\cal C}^{'} \in S({\cal C}_0 )}{\cal C}^{'} \neq \emptyset$,
where $S({\cal C}_0) = \{ {\cal C}^{'} \subseteq {\cal C} \ | \ {\sf desc}({\cal C}^{'}) = {\sf desc}({\cal C}_0) = R \}$.
We prove this theorem in three steps.

(1)   ${\cal C}_{L} \in S({\cal C}_0)$, that is ${\sf desc}({\cal C}_{L}) = R$.
For any $1 \leq j \leq n$, we consider the following cases.

(1.1)   $R(j) = \{1\}$. For any
${\bf c} \in {\cal C}_{L} $, ${\bf c}(j)=1$ according to the processes deriving ${\cal C}_{L}$.
So, ${\cal C}_{L}(j) = \{1\} = R(j)$.

(1.2)   $R(j) = \{0\}$. For any ${\bf c} \in {\cal C}_{L}$, ${\bf c}(j)=0$
according to the processes  deriving ${\cal C}_{L}$.
So, ${\cal C}_{L}(j) = \{0\} = R(j)$.

(1.3)  $R(j) = \{0, 1\}$.  Since ${\sf desc}({\cal C}_{0}) = R$,
we know that there exist ${\bf c}_1, {\bf c}_2 \in {\cal C}_0 \subseteq {\cal C}_{L}$
such that ${\bf c}_1(j) =0$ and ${\bf c}_2(j) =1$, respectively.
This implies ${\cal C}_{L}(j) = \{0, 1\} = R(j)$.

According to (1.1)-(1.3) above, for any $1 \leq j \leq n$, we have
${\cal C}_{L}(j)= R(j)$,
which implies ${\sf desc}({\cal C}_{L}) = R$.

(2)   We want to  show that  for any
${\bf x} \in \bigcap_{{\cal C}^{'} \in S({\cal C}_0 )}{\cal C}^{'}$,
there exists $1 \leq j \leq n$, such that  ${\bf x}(j) =1$ and ${\bf c}(j) = 0$
for any ${\bf c} \in {\cal C}_{L} \setminus \{ {\bf x}\}$,
or ${\bf x}(j) =0$ and ${\bf c}(j) = 1$ for any ${\bf c} \in {\cal C}_{L} \setminus \{ {\bf x}\}$.
If this is not true, then for any  $1 \leq j \leq n$,  ${\bf x}(j) =1$ implies that there exists
${\bf c}_1 \in {\cal C}_{L} \setminus \{ {\bf x} \}$ such that ${\bf c}_1(j) = 1$,
and ${\bf x}(j) =0$ implies that there exists
${\bf c}_2\in {\cal C}_{L} \setminus \{ {\bf x} \}$ such that ${\bf c}_2(j) = 0$.
Then we have  ${\sf desc}({\cal C}_{L}) = {\sf desc}({\cal C}_{L}\setminus\{ {\bf x} \})$.
Since ${\cal C}_{L} \in S({\cal C}_0)$ by (1),
we can have  ${\cal C}_{L} \setminus \{ {\bf x}\} \in S({\cal C}_0)$, and
${\bf x} \notin \bigcap_{{ \cal C}^{'} \in S({\cal C}_0)} {\cal C}^{'}$, a contradiction.

(3)   At last, according to Algorithm \ref{AlgSSC} and (2), it suffices to show that any user $u$
assigned with a codeword ${\bf x} \in \bigcap_{{\cal C}^{'} \in S({\cal C}_0)}{\cal C}^{'}$
is a colluder. Assume that $u$ is not a colluder.
Then for any ${\cal C}^{'} \in S({\cal C}_0)$,
we have ${\cal C}^{'} \setminus\{{\bf x}\} \in S({\cal C}_0)$,
which implies ${\bf x} \notin \bigcap_{{\cal C}^{'}  \in S({\cal C}_0)}{\cal C}^{'}$,
a contradiction.




The proof is then completed.
\end{IEEEproof}

The following is a construction for binary $t$-SMIPPCs
from $q$-ary $t$-SMIPPCs, which makes the research of $q$-ary $t$-SMIPPCs interesting.

\begin{lemma}
\label{compoconstru}
If there exists a $t$-SMIPPC$(n,M,q)$, then there exists a $t$-SMIPPC$(nq,M,2)$.
\end{lemma}
\begin{IEEEproof}
Let ${\cal C} = \{{\bf c}_1, {\bf c}_2, \ldots, {\bf c}_M\}$ be a $t$-SMIPPC$(n, M, q)$ defined on $Q = \{0,1, \ldots, q-1\}$,
and ${\cal E} = \{{\bf e}_1, {\bf e}_2, \ldots, {\bf e}_{q}\}$, where ${\bf e}_i$ is the $i$-th column identity vector,
i.e., all its coordinates are $0$ except the $i$-th one being $1$.
Let $f: Q \longrightarrow {\cal E}$ be the bijective mapping such that $f(i) = {\bf e}_{i+1}$.
For any codeword ${\bf c} = ({\bf c}(1), {\bf c}(2), \ldots, {\bf c}(n))^{T} \in {\cal C}$,
we define $f({\bf c}) = (f({\bf c}(1)), f({\bf c}(2)), \ldots, f({\bf c}(n)))^{T}$.
Obviously, $f({\bf c})$ is a binary column vector of length $nq$.
We define a new $(nq, M, 2)$ code ${\cal F} = \{f({\bf c}_1), f({\bf c}_2), \ldots, f({\bf c}_M)\}$,
and show that ${\cal F}$ is in fact a $t$-SMIPPC.

Consider any ${\cal F}_0 \subseteq {\cal F}$ with $|{\cal F}_0| \le t$,
and $S({\cal F}_0) = \{ {\cal F}^{'} \subseteq {\cal F} \ | \ {\sf desc}({\cal F}^{'}) = {\sf desc}({\cal F}_0) \}
= \{ {\cal F}_0, {\cal F}_1, \ldots, {\cal F}_r\}$.
Each ${\cal F}_i$ corresponds to a subcode ${\cal C}_i \subseteq {\cal C}$ such that $|{\cal C}_i|= |{\cal F}_i|$,
where ${\cal F}_i = \{f({\bf c}) \ | \ {\bf c} \in {\cal C}_i\}$.
Since ${\sf desc}({\cal F}_0) = {\sf desc}({\cal F}_1) = \cdots = {\sf desc}({\cal F}_r)$,
we immediately have ${\sf desc}({\cal C}_0) = {\sf desc}({\cal C}_1) = \cdots = {\sf desc}({\cal C}_r)$.
Since ${\cal C}$ is a $t$-SMIPPC$(n, M, q)$ and $|{\cal C}_0|= |{\cal F}_0| \leq t$, we have $\bigcap_{i = 0}^{r}{\cal C}_i \neq \emptyset$.
Let ${\bf c} \in \bigcap_{i = 0}^{r}{\cal C}_i$, then ${\bf c} \in {\cal C}_i$ for any $0 \leq i \leq r$,
which implies $f({\bf c}) \in {\cal F}_i$ for any $0 \leq i \leq r$, and thus $f({\bf c}) \in \bigcap_{i = 0}^{r}{\cal F}_i$.
Therefore, $\bigcap_{i = 0}^{r}{\cal F}_i \neq \emptyset$. This completes the proof.
\end{IEEEproof}

\section{Optimal $t$-SMIPPC$(2, M, q)$s with small $t$}   %
\label{optimalSMIPPC}

Let $M_{SMIPPC}(t,n,q) = \mbox{max} \{M \ | $\  there exists a $t$-SMIPPC$(n,M,q)\}$.
A $t$-SMIPPC$(n, M,q)$ is said to be optimal if $M = M_{SMIPPC}(t,n,q)$.
Similarly, we can define $M_{MIPPC}(t,n,q) = \mbox{max} \{M \ | \ \mbox{there exists a}  \ t \mbox{-MIPPC}(n, \\ M,q)\}$
and optimal $t$-MIPPC$(n, M, q)$s. In this section, We establish  two equivalences
in Corollary \ref{equi2} and Theorem \ref{equi7}, respectively.
Based on these two relationships and the known results in Lemmas \ref{resulSC} and \ref{optimal},
several infinite series of optimal $t$-SMIPPC$(2, M, q)$s with $t=2, 3$ are derived.

Separable codes ($\overline{t}$-SCs) defined as follows were well studied in \cite{CFJLM1, CJM, CM}.

\begin{definition} $(${\rm \cite{CM}}$)$
\label{defSC}
Let ${\cal C}$ be an $(n,M,q)$ code and $t \ge 2$ be an integer.
${\cal C}$ is a $\overline{t}$-separable code, or $\overline{t}$-SC$(n,M,q)$,
if for any ${\cal C}_1, {\cal C}_2 \subseteq {\cal C}$ such that
$1 \le |{\cal C}_1| \le t$, $1 \le |{\cal C}_2| \le t$ and ${\cal C}_1 \neq {\cal C}_2$,
we have ${\sf desc}({\cal C}_1) \neq {\sf desc}({\cal C}_2)$.
\end{definition}

Similar to $M_{SMIPPC}(t,n,q)$ and optimal $t$-SMIPPC$(n$, $M,q)$s,
we can define $M_{SC}(\overline{t},n,q) = \mbox{max} \{M \ | \ \mbox{there} \\ \mbox{exists a}  \ \overline{t} \mbox{-SC}(n,M,q)\}$
 and optimal $\overline{t}$-SC$(n, M, q)$s.
The following optimal SCs come from \cite{CFJLM1, CJM}.

\begin{lemma} $(${\rm \cite{CFJLM1, CJM}}$)$
\label{resulSC}
Let $k \geq 2$ be a prime power.
Then there is an optimal  $\overline{2}$-SC$(2,M,q)$ for any $q \in \{ k^2 -1, k^2+k-2, k^2+k-1, k^2+k, k^2+k+1\}$.
\end{lemma}

The following result  will be used to obtain  the equivalence between a $2$-SMIPPC$(2,M,q)$ and a $\overline{2}$-SC$(2,M,q)$.

\begin{theorem} {\rm (\cite{CFJLM})}
\label{equi6}
Let $\cal C$ be an $(n,M,q)$ code.
Then $\cal C$ is a $2$-MIPPC$(n,M,q)$ if and only if it is a $\overline{2}$-SC$(n,M,q)$.
\end{theorem}

\begin{theorem}
\label{equi}
Let $\cal C$ be a $(2,M,q)$ code.
Then $\cal C$ is a $2$-SMIPPC$(2,M,q)$ if and only if it is a $2$-MIPPC$(2,M,q)$.
\end{theorem}
\begin{IEEEproof} According to  Lemma \ref{rela5}, it suffices to consider the sufficiency.
Let ${\cal C}$ be a $2$-MIPPC$(2, M, q)$, which implies that ${\cal C}$ is a
$\overline{2}$-SC$(2, M, q)$ from Theorem \ref{equi6}.
Assume that ${\cal C}$ is not a $2$-SMIPPC$(2, M, q)$.
Then there exists ${\cal C}_0 \subseteq {\cal C}$, $1 \leq |{\cal C}_0| \leq 2$,
such that $\bigcap_{{\cal C}^{'} \in S({\cal C}_0)}{\cal C}^{'} =  \emptyset$,
where $S({\cal C}_0) = \{ {\cal C}^{'} \subseteq {\cal C} \ | \ {\sf desc}({\cal C}^{'}) = {\sf desc}({\cal C}_0)\}$.
If $|{\cal C}_0|=1$, then it is clear that $S({\cal C}_0) =  \{ {\cal C}_0\}$,
which implies $\bigcap_{{\cal C}^{'} \in S({\cal C}_0)}{\cal C}^{'} = {\cal C}_0 \neq \emptyset$, a contradiction.
So $|{\cal C}_0|=2$. Let ${\cal C}_0 = \{ {\bf c}_1, {\bf c}_2 \}$, ${\bf c}_i = (a_i, b_i)^{T}$, where $i = 1, 2$.
Obviously, for any ${\cal C}^{'} \in S({\cal C}_0)$,
we have ${\cal C}^{'} \subseteq {\sf desc}({\cal C}_0) \bigcap {\cal C}$.
We now consider the Hamming distance $d({\bf c}_1, {\bf c}_2)$ of ${\bf c}_1$ and ${\bf c}_2$.

(1)  If $d({\bf c}_1, {\bf c}_2) = 1$, we may assume $a_1 = a_2$, $b_1 \neq b_2$. We can easily see
that $S({\cal C}_0) = \{ {\cal C}_0\}$, which implies $\bigcap_{{\cal C}^{'} \in S({\cal C}_0)}{\cal C}^{'}
= {\cal C}_0 \neq \emptyset$, a contradiction. So this case is impossible.

(2)  If $d({\bf c}_1, {\bf c}_2) = 2$, then $a_1 \neq a_2$, $b_1 \neq b_2$,
and ${\sf desc}({\cal C}_0) = \{ {\bf c}_1, {\bf c}_2, {\bf c}_3, {\bf c}_4 \}$,
where ${\bf c}_3 = (a_1, b_2)^{T}$ and ${\bf c}_4 = (a_2, b_1)^{T}$.
Then $|{\sf desc}({\cal C}_0) \bigcap {\cal C}| \leq 3$.
Otherwise, if $|{\sf desc}({\cal C}_0) \bigcap {\cal C}| = 4$, i.e.,
${\sf desc}({\cal C}_0) \bigcap {\cal C} = \{ {\bf c}_1, {\bf c}_2, {\bf c}_3, {\bf c}_4 \}$,
then ${\sf desc}(\{ {\bf c}_1, {\bf c}_2\}) = {\sf desc}(\{ {\bf c}_3, {\bf c}_4\})$,
a contradiction to the fact that ${\cal C}$ is a $\overline{2}$-{\rm SC}.
Since ${\cal C}$ is a $\overline{2}$-{\rm SC}$(2, M, q)$,
for any ${\cal C}^{'} \in S({\cal C}_0)$, ${\cal C}^{'} \neq {\cal C}_0$,
we have $|{\cal C}^{'}| \geq 3$. Together with the facts ${\cal C}^{'} \subseteq {\sf desc}({\cal C}_0) \bigcap {\cal C}$
and $|{\sf desc}({\cal C}_0) \bigcap {\cal C}| \leq 3$,
one can derive ${\cal C}^{'} = {\sf desc}({\cal C}_0) \bigcap {\cal C}$.
Hence ${\cal C}_0 \subseteq {\sf desc}({\cal C}_0) \bigcap {\cal C} = {\cal C}^{'}$,
which implies $\bigcap_{{\cal C}^{'} \in S({\cal C}_0)}{\cal C}^{'} = {\cal C}_0 \neq \emptyset$, a contradiction.
So this case is impossible.

The proof is then completed.
\end{IEEEproof}

The following result comes from Theorems  \ref{equi6} and \ref{equi}.

\begin{corollary}
\label{equi2}
Let $\cal C$ be an $(n,M,q)$ code.
Then $\cal C$ is a $2$-SMIPPC$(2,M,q)$ if and only if it is a $\overline{2}$-SC$(2,M,q)$.
\end{corollary}
F

Thus, according to Lemma \ref{resulSC} and Corollary \ref{equi2}, one can obtain optimal $2$-SMIPPC$(2,M,q)$s.

\begin{corollary}
\label{resulSMIPPC}
Let $k \geq 2$ be a prime power.
Then there is an optimal  $2$-SMIPPC$(2,M,q)$ for any $q \in \{ k^2 -1, k^2+k-2, k^2+k-1, k^2+k, k^2+k+1\}$.
\end{corollary}

Similarly, we also find an equivalence between a $3$-SMIPPC$(2, M, q)$
and a $3$-MIPPC$(2, M, q)$ as follows.

\begin{theorem}
\label{equi7}
Let ${\cal C}$ be an $(2, M, q)$ code. Then ${\cal C}$ is a $3$-SMIPPC$(2, M, q)$
 if and only if it is a $3$-MIPPC$(2, M, q)$.
\end{theorem}
\begin{IEEEproof}  By Lemma \ref{rela5}, it suffices to consider the sufficiency.
Suppose that ${\cal C}$ is a $3$-MIPPC$(2, M, q)$.
Then ${\cal C}$ is also a $2$-MIPPC$(2, M, q)$, which implies that ${\cal C}$ is a
$2$-SMIPPC$(2, M, q)$ from Theorem \ref{equi}.
Assume that  ${\cal C}$ is not a $3$-SMIPPC$(2, M, q)$.
Then there exists ${\cal C}_0 \subseteq {\cal C}$, $1 \leq |{\cal C}_0| \leq 3$,
such that $\bigcap_{{\cal C}^{'} \in S({\cal C}_0)}{\cal C}^{'} =  \emptyset$,
where $S({\cal C}_0) = \{ {\cal C}^{'} \subseteq {\cal C} \ | \ {\sf desc}({\cal C}^{'}) = {\sf desc}({\cal C}_0)\}$.
Obviously, for any ${\cal C}^{'} \in S({\cal C}_0)$,
we have ${\cal C}^{'} \subseteq {\sf desc}({\cal C}_0) \bigcap {\cal C}$.
Then, at least one of the following cases should occur. However, we can
prove none of them is possible.

(1)    $1 \leq |{\cal C}_0| \leq 2$.
Since ${\cal C}$ is a $2$-SMIPPC$(2, M, q)$, $\bigcap_{{\cal C}^{'} \in S({\cal C}_0)}{\cal C}^{'} \neq \emptyset$,
a contradiction. So this case is impossible.

(2)   If $|{\cal C}_0| = 3$, then let ${\cal C}_0 = \{ {\bf c}_1, {\bf c}_2, {\bf c}_3\}$,
where ${\bf c}_i = (a_i, b_i)^{T}$, $1 \leq i \leq 3$.

(2.1)   If $a_1 = a_2 = a_3$, then $b_i \neq b_j$, $ 1 \leq i < j \leq 3$. We can easily see that $S({\cal C}_0) = \{ {\cal C}_0\}$,
which implies $\bigcap_{{\cal C}^{'} \in S({\cal C}_0)}{\cal C}^{'} = {\cal C}_0 \neq \emptyset$, a contradiction. So this case is impossible.

(2.2)   If $a_1 = a_2 \neq a_3$, then $b_1 \neq b_2$.
Let ${\cal C}_1 = ({\sf desc}({\cal C}_0) \bigcap {\cal C}) \setminus {\cal C}_0$.
Then $b_1 \notin {\cal C}_1(2)$ or $b_2 \notin {\cal C}_1(2)$.
Otherwise, $b_1, b_2 \in {\cal C}_1(2)$, which implies that $(a_3, b_1)^{T}, (a_3, b_2)^{T} \in {\cal C}$.
Then we  have ${\sf desc}( \{ {\bf c}_1, (a_3, b_2)^{T} \}) = {\sf desc}(\{ {\bf c}_2, (a_3, b_1)^{T} \})$,
and $\{ {\bf c}_1, (a_3, b_2)^{T} \} \bigcap \{ {\bf c}_2, (a_3, b_1)^{T} \} = \emptyset$, a contradiction to
the definition of a $3$-MIPPC.

(2.2.A)   If $b_1 \notin {\cal C}_1(2)$, then ${\bf c}_1$ is the only codeword such that
${\bf c}_1 \in {\sf desc}({\cal C}_0) \bigcap {\cal C}$ and ${\bf c}_1(2) = b_1$.
Since ${\cal C}^{'} \subseteq {\sf desc}({\cal C}_0) \bigcap {\cal C}$,
we should have ${\bf c}_1 \in {\cal C}^{'}$ for any ${\cal C}^{'} \in S({\cal C}_0)$.
Otherwise, if ${\bf c}_1 \notin {\cal C}^{'}$,  then $b_1 \notin {\cal C}^{'}(2)$, which implies
${\sf desc}({\cal C}^{'}) \neq {\sf desc}({\cal C}_0)$ as $b_1 \in {\cal C}_0(2)$, a contradiction.
So, in this case, $ {\bf c}_1 \in \bigcap_{{\cal C}^{'} \in S({\cal C}_0)}{\cal C}^{'}$,
which implies $\bigcap_{{\cal C}^{'} \in S({\cal C}_0)}{\cal C}^{'} \neq \emptyset$, a contradiction to the assumption.
So this case is impossible.

(2.2.B)    If $b_2 \notin {\cal C}_1(2)$, similar to (2.2.A),
we can have $ {\bf c}_2 \in \bigcap_{{\cal C}^{'} \in S({\cal C}_0)}{\cal C}^{'}$,
which implies $\bigcap_{{\cal C}^{'} \in S({\cal C}_0)}{\cal C}^{'} \neq \emptyset$, a contradiction to the assumption.
So this case is impossible.

(2.3)   If $a_i \neq a_j$, $1 \leq i < j \leq 3$,  we only need to consider the case $b_i \neq b_j$ , $1 \leq i < j \leq 3$,
since we can consider the other two cases in a similar way with (2.1) and (2.2). In this case, we have\\
\indent \ \ \ \  \ \ \ \  \ \ \ \  \ \ \ \ ${\bf c}_1 \ \ \   {\bf c}_2 \ \ \  {\bf c}_3 \ \ \ {\bf c}_4 \ \ \  {\bf c}_5 \ \ \ {\bf c}_6 \ \ \   {\bf c}_7 \ \ \ {\bf c}_8 \ \ \ {\bf c}_9$
$$ {\sf desc}({\cal C}_0)=
\left(
  \begin{array}{ccc|cccccc}
    a_1 & a_2 & a_3 & a_1 & a_1 & a_2 & a_2 & a_3 & a_3\\
    b_1 & b_2 & b_3 & b_2 & b_3 & b_1 & b_3 & b_1 & b_2\\
  \end{array}
\right)
$$
If ${\sf desc}({\cal C}_0) \bigcap {\cal C} = {\cal C}_0$, we can check that for any ${\cal C}^{'}\in S({\cal C}_0)$,
${\cal C}^{'} \subseteq  {\sf desc}({\cal C}^{'}) \bigcap {\cal C} = {\sf desc}({\cal C}_0) \bigcap {\cal C} ={\cal C}_0$,
then $| {\cal C}^{'} | \leq | {\cal C}_{0} | = 3$,
and hence $\bigcap_{{\cal C}^{'} \in S({\cal C}_0)}{\cal C}^{'} \neq \emptyset$ since
${\cal C}$ is a $3$-MIPPC,  a contradiction to the assumption.
So ${\sf desc}({\cal C}_0) \bigcap {\cal C} $ contains at least one of the words
${\bf c}_4, {\bf c}_5, {\bf c}_6, {\bf c}_7, {\bf c}_8, {\bf c}_9$.
Without loss of generality, we only need  to consider the case $ {\bf c}_4 \in {\sf desc}({\cal C}_0) \bigcap {\cal C}$.
Then ${\bf c}_6 \notin {\sf desc}({\cal C}_0) \bigcap {\cal C}$, otherwise,
${\sf desc}( \{ {\bf c}_1, {\bf c}_2\}) = {\sf desc}(\{ {\bf c}_4, {\bf c}_6\})$,
and $\{ {\bf c}_1, {\bf c}_2\} \bigcap \{ {\bf c}_4, {\bf c}_6\} = \emptyset$,
a contradiction to the definition of a $3$-MIPPC.
We will show that  ${\bf c}_7, {\bf c}_8 \in {\sf desc}({\cal C}_0) \bigcap {\cal C}$.
If ${\bf c}_7 \notin {\sf desc} ({\cal C}_0) \bigcap {\cal C}$ (or ${\bf c}_8 \notin {\sf desc}({\cal C}_0) \bigcap {\cal C})$,
then for any ${\cal C}^{'} \in S({\cal C}_0)$, we have ${\bf c}_2 \in {\cal C}^{'}$ (or ${\bf c}_1 \in {\cal C}^{'}$), otherwise,
$a_2 \notin {\cal C}^{'}(1)$ (or $b_1 \notin {\cal C}^{'}(2)$), which implies ${\sf desc}({\cal C}^{'}) \neq {\sf desc}({\cal C}_0)$
since $a_2 \in {\cal C}_0(1)$ (or $b_1 \in {\cal C}_0(2)$).
Hence $ {\bf c}_2 \in \bigcap_{{\cal C}^{'} \in S ({\cal C}_0)}{\cal C}^{'}$
(or $ {\bf c}_1 \in \bigcap_{{\cal C}^{'} \in S({\cal C}_0)}{\cal C}^{'}$),
which implies  $\bigcap_{{\cal C}^{'} \in S({\cal C}_0)}{\cal C}^{'} \neq \emptyset$,
a contradiction to the assumption.
So, ${\bf c}_4, {\bf c}_7, {\bf c}_8 \in {\sf desc}({\cal C}_0) \bigcap {\cal C}$.
Then ${\sf desc}( \{ {\bf c}_1, {\bf c}_2, {\bf c}_3\}) = {\sf desc}(\{{\bf c}_4, {\bf c}_7, {\bf c}_8\})$,
while $\{ {\bf c}_1, {\bf c}_2, {\bf c}_3\} \bigcap \{ {\bf c}_4, {\bf c}_7, {\bf c}_8\} = \emptyset$,
a contradiction to the definition of a $3$-MIPPC.
So this case is impossible.

The proof is then completed.
\end{IEEEproof}

The above theorem shows that the following optimal $3$-MIPPCs of length $2$ are in fact optimal $3$-SMIPPCs of length $2$.

\begin{lemma}  $(${\rm \cite{CFJLM}}$)$
\label{optimal}
There exists an optimal $3$-MIPPC$(2$, $(k^2+1)(k+1)^2, (k^2+1)(k+1))$ for any prime power $k$.
\end{lemma}

\begin{corollary}
\label{resul}
There exists an optimal $3$-SMIPPC$(2, (k^2+1)(k+1)^2, (k^2+1)(k+1))$ for any prime power $k$.
\end{corollary}

\section{$2$-SMIPPC$(3, M, q)$}   %
\label{construction}

In this section, we will investigate the combinatorial properties of a $2$-SMIPPC$(3,M,q)$,
and then derive forbidden configurations of a $2$-SMIPPC$(3, M, q)$.
Optimal $2$-SMIPPC$(3, M, q)$s are also constructed for each $q \equiv 0, 1, 2, 5 \pmod{6}$.

\subsection{General idea}   %

At first, one can easily derive the following result from Lemma \ref{rela5} and Theorem \ref{equi6}.

\begin{corollary}
\label{rela6}
Any $2$-SMIPPC$(n,M,q)$ is a $\overline{2}$-SC$(n,M,q)$.
\end{corollary}

Thus, $M_{SMIPPC}(2,n,q) \leq M_{SC}(\overline{2},n,q)$, and we can
investigate $2$-SMIPPC$(3, M, q)$s based on $\overline{2}$-SC$(3, M, q)$s.

\begin{lemma}{\rm (\cite{CJM})}
\label{SC}
A $(3, M, q)$ code is a $\overline{2}$-{\rm SC}$(3, M, q)$ on $Q$ if and only if
$|{\cal A}_{g_1}^{j} \bigcap {\cal A}_{g_2}^{j}| \leq 1$ holds for any
positive integers $1 \leq j \leq 3$, and any distinct $g_1, g_2 \in Q$.
\end{lemma}

\begin{lemma}{\rm (\cite{CJM})}
\label{upperbound}
For any $\overline{2}$-SC$(3,M,q)$, we have $M \leq q^2 + \frac{q(q-1)}{2}$.
\end{lemma}

Then an upper bound on the size of a $2$-SMIPPC$(3,M,q)$ can be derived by Corollary \ref{rela6} and Lemma \ref{upperbound}.

\begin{theorem}
\label{upperbound1}
For any $2$-SMIPPC$(3,M,q)$, we have $M \leq q^2 + \frac{q(q-1)}{2}$.
\end{theorem}

Next, we try to find out forbidden configurations of a $2$-SMIPPC$(3, M, q)$.

\begin{lemma} {\rm (\cite{JCM})}
\label{forbiconfi2}
Let ${\cal C}$ be a $\overline{2}$-{\rm SC}$(3, M, q)$.
If there exist ${\cal C}_0, {\cal C}^{'} \subseteq {\cal C}$, $|{\cal C}_0| \leq 2$, such that
${\sf desc}({\cal C}_0) = {\sf desc}({\cal C}^{'})$ and ${\cal C}_0 \not\subseteq  {\cal C}^{'}$, then
${\sf desc}({\cal C}_0)\bigcap {\cal C}$ is of one of the following four types:
\begin{eqnarray*}
\begin{array}{cccc}
\hbox{Type {\bf I}:} & \hbox{Type {\bf II}:}  \\
\left(
  \begin{array}{cc|cccccc}
    a_1 & a_2 & a_1 & a_1 \\
    b_1 & b_2 & b_1 & b_2 \\
    e_1 & e_2 & e_2 & e_1 \\
  \end{array}
\right),&
\left(
  \begin{array}{cc|cccccc}
    a_1 & a_2 & a_1 & a_2 \\
    b_1 & b_2 & b_1 & b_1 \\
    e_1 & e_2 & e_2 & e_1 \\
  \end{array}
\right),\\[1cm]
\hbox{Type {\bf III:} } & \hbox{Type {\bf IV:}}\\
\left(
  \begin{array}{cc|cccccc}
    a_1 & a_2 & a_1 & a_2 \\
    b_1 & b_2 & b_2 & b_1 \\
    e_1 & e_2 & e_1 & e_1 \\
  \end{array}
\right),&
\left(
  \begin{array}{cc|cccccc}
    a_1 & a_2 & a_1 & a_1 & a_2\\
    b_1 & b_2 & b_1 & b_2 & b_1\\
    e_1 & e_2 & e_2 & e_1 & e_1\\
  \end{array}
\right),
\end{array}
\end{eqnarray*}
where ${\cal C}_0 = \{ {\bf c}_1, {\bf c}_2 \}$, ${\bf c}_i = (a_i, b_i, e_i)$, $i = 1, 2$,
and $a_1 \neq a_2$, $b_1 \neq b_2$, $e_1 \neq e_2$.
\end{lemma}

\begin{theorem}
\label{forbiconfi3}
Let ${\cal C}$ be a $\overline{2}$-{\rm SC}$(3, M, q)$. Then ${\cal C}$ is a $2$-SMIPPC$(3, M, q)$
if and only if for any ${\cal C}_0 = \{ {\bf c}_1, {\bf c}_2 \} =  \{(a_1, b_1, e_1)^{T}, (a_2, b_2, e_2)^{T}\} \subseteq {\cal C}$,
where $a_1 \neq a_2$, $b_1 \neq b_2$, and $e_1 \neq e_2$,
we have ${\sf desc}({\cal C}_0)\bigcap {\cal C}$ is not of type  {\bf IV} of Lemma {\rm \ref{forbiconfi2}}:
$$
\left(
  \begin{array}{cc|cccccc}
    a_1 & a_2 & a_1 & a_1 & a_2\\
    b_1 & b_2 & b_1 & b_2 & b_1\\
    e_1 & e_2 & e_2 & e_1 & e_1\\
  \end{array}
\right)
$$
\end{theorem}
\begin{IEEEproof}  Suppose that ${\cal C}$ is a $2$-SMIPPC$(3, M, q)$.
If there exists ${\cal C}_0 = \{ {\bf c}_1, {\bf c}_2 \} =  \{(a_1, b_1, e_1)^{T}, (a_2, b_2, e_2)^{T}\}  \subseteq {\cal C}$, where
$a_1 \neq a_2$, $b_1 \neq b_2$, and $e_1 \neq e_2$, such that
${\sf desc}({\cal C}_0)\bigcap {\cal C}$ is  of  type  {\bf IV}, then we can derive that
${\sf desc}( \{ {\bf c}_1, {\bf c}_2\}) = {\sf desc}(\{ (a_1, b_1, e_2)^{T}, (a_1, b_2, e_1)^{T}, (a_2, b_1, e_1)^{T}\})$,
while $\{ {\bf c}_1, {\bf c}_2\} \bigcap \{ (a_1, b_1, e_2)^{T}, (a_1, b_2, e_1)^{T}, \\ (a_2, b_1, e_1)^{T}\} = \emptyset$,
a contradiction to the definition of a $2$-SMIPPC.

Conversely, suppose that ${\cal C}$ is a $\overline{2}$-{\rm SC}$(3, M, q)$,
and for any ${\cal C}_0 = \{ {\bf c}_1, {\bf c}_2 \} =  \{(a_1, b_1, e_1)^{T}, (a_2, b_2, e_2)^{T}\} $
$ \subseteq {\cal C}$, where
$a_1 \neq a_2$, $b_1 \neq b_2$, and $e_1 \neq e_2$,
we have ${\sf desc}({\cal C}_0)\bigcap {\cal C}$ is not of  type {\bf IV}.
We will show $\bigcap_{{\cal C}^{'} \in S({\cal C}_0)}{\cal C}^{'}  \neq \emptyset$

(1)   If for any ${\cal C}^{'} \in S({\cal C}_0)$, we have ${\cal C}_0 \subseteq {\cal C}^{'}$,
then $\bigcap_{{\cal C}^{'} \in S({\cal C}_0)}{\cal C}^{'} = {\cal C}_0 \neq \emptyset$.

(2)   If there exists ${\cal C}^{''} \in S({\cal C}_0)$ such that ${\cal C}_0 \not\subseteq {\cal C}^{''}$,
then by Lemma \ref{forbiconfi2}, we know that ${\sf desc}({\cal C}_0)\bigcap {\cal C}$ is of one of the
four types mentioned in Lemma \ref{forbiconfi2}.
Since  ${\sf desc}({\cal C}_0)\bigcap {\cal C}$ is not of
type {\bf IV}, we know that ${\sf desc}({\cal C}_0)\bigcap {\cal C}$ is of one of the
types {\bf I}, {\bf II}, {\bf III}.

(2.1)   If ${\sf desc}({\cal C}_0)\bigcap {\cal C}$ is of type {\bf I}, then for any ${\cal C}^{'} \in S({\cal C}_0)$,
we have ${\cal C}^{'} \subseteq {\sf desc}({\cal C}_0) \bigcap {\cal C}$, and thus
${\bf c}_2 \in {\cal C}^{'}$, otherwise, $a_2 \notin {\cal C}^{'}(1)$,
which implies ${\sf desc}({\cal C}^{'}) \neq {\sf desc}({\cal C}_0)$, a contradiction.
So we have ${\bf c}_2 \in \bigcap_{{\cal C}^{'} \in S({\cal C}_0)}{\cal C}^{'}$,
which implies $\bigcap_{{\cal C}^{'} \in S({\cal C}_0)}{\cal C}^{'} \neq \emptyset$.

(2.2)   If ${\sf desc}({\cal C}_0)\bigcap {\cal C}$ is of type {\bf II}, then for any ${\cal C}^{'} \in S({\cal C}_0)$,
we have ${\cal C}^{'} \subseteq  {\sf desc}({\cal C}_0) \bigcap {\cal C}$,
and thus ${\bf c}_2 \in {\cal C}^{'}$, otherwise, $b_2 \notin {\cal C}^{'}(2)$,
which implies ${\sf desc}({\cal C}^{'}) \neq {\sf desc}({\cal C}_0)$, a contradiction.
So we have ${\bf c}_2 \in \bigcap_{{\cal C}^{'} \in S({\cal C}_0)}{\cal C}^{'}$,
which implies $\bigcap_{{\cal C}^{'} \in S({\cal C}_0)}{\cal C}^{'} \neq \emptyset$.

(2.3)   If ${\sf desc}({\cal C}_0)\bigcap {\cal C}$ is of type {\bf III}, then for any ${\cal C}^{'} \in S({\cal C}_0)$,
we have ${\cal C}^{'} \subseteq {\sf desc}({\cal C}_0) \bigcap {\cal C}$,
and thus ${\bf c}_2 \in {\cal C}^{'}$, otherwise, $e_2 \notin {\cal C}^{'}(3)$,
which implies ${\sf desc}({\cal C}^{'}) \neq {\sf desc}({\cal C}_0)$, a contradiction.
So we have ${\bf c}_2 \in \bigcap_{{\cal C}^{'} \in S({\cal C}_0)}{\cal C}^{'}$,
which implies $\bigcap_{{\cal C}^{'} \in S({\cal C}_0)}{\cal C}^{'} \neq \emptyset$.

Therefore, ${\cal C}$ is a $2$-SMIPPC$(3, M, q)$.
 \end{IEEEproof}

To construct $2$-SMIPPC$(3,M,q)$s, a cyclic difference matrix is needed.

\begin{definition}
\label{defCDM}
A cyclic difference matrix $(q, k, 1)$-CDM is a $k \times q$ matrix $D = (d_{ij})$ with $d_{ij} \in Z_q$
such that for any $1 \leq i_1 \neq i_2 \leq k$, the differences $d_{i_1j} - d_{i_2j}$, $1 \leq j \leq q$,
comprise all the elements of $Z_q$.
\end{definition}

Similar to \cite{CJM}, suppose that  there exists a $(q, 3, 1)$-CDM  $D$. Without loss of generality, we may assume that
$$ \indent \indent\indent\indent \indent
 D =
\left(
  \begin{array}{cccc}
     0   &  0  &  \cdots  &     0   \\
     0   &  1  &  \cdots  &    q-1  \\
     x_0   &  x_1  &  \cdots  &   x_{q-1} \\
  \end{array}
\right). \ \ \  {\rm (i)}
$$
Let $S$ be a $3 \times w$ matrix on $Z_q$ as follows.
$$\indent \indent\indent\indent \indent
S =
\left(
  \begin{array}{cccc}
     0      &    0   &   \cdots    &   0   \\
     s_1    &   s_2  &    \cdots   &  s_{w}  \\
     t_1    &  t_2   &  \cdots     & t_{w} \\
  \end{array}
\right).  \ \ \ \ \  {\rm (ii)}
$$
Let
\begin{eqnarray*}&&{\cal C}_D = \{ {\bf c} + g \ | \ {\bf c} \in D, g \in Z_{q} \}, \\
&&{\cal C}_S = \{ {\bf c} + g \ | \ {\bf c} \in S, g \in Z_{q} \}, \
{\cal C} = {\cal C}_D \bigcup {\cal C}_S. \ \ {\rm (iii)}\end{eqnarray*}

\begin{theorem}{\rm (\cite{CJM})}
\label{equi8}
Suppose that $D$ is a $(q, 3, 1)$-CDM in the form {\rm (i)} and $S$ is a $3 \times w$ matrix in the form {\rm (ii)}, where
$|\{s_1, s_2, \ldots, s_w\}| = |\{t_1, t_2, \ldots, t_w\}| = |\{t_1-s_1, t_2-s_2, \ldots, t_w-s_w\}| = w$. Then, the following
two statements are equivalent:
\begin{itemize}
\item[(1)] ${\cal C}$ in the form {\rm (iii)} is a $\overline{2}$-SC$(3, q(q+w), q)$;
\item[(2)]  For any two columns $(0, s_i, t_i)^{T}$ and $(0, s_j, t_j)^{T}$ in $S$, $1 \leq i \neq j \leq w$,
suppose $(0, y, x_y)^{T}$, $(0, z, x_z)^{T}$, $(0, y_i, x_{y_i})^{T}$, $(0, y_j, x_{y_j})^{T}$,
$(0, z_i, x_{z_i})^{T}$, $(0, z_j, x_{z_j})^{T} \in D$, where $y, z, y_i, y_j, z_i, z_j \in Z_q$, such that
$$
\left\{\begin{array}{rl}
t_i - s_i = x_y - y,   \\[2pt]
t_j - s_j = x_z - z,   \\[2pt]
t_i  = x_{y_i},  \ \ \ \ \ \ \ \ \ \   \\[2pt]
t_j  = x_{y_j},  \ \ \ \ \ \ \ \ \ \ \\[2pt]
s_i  = z_i, \ \ \ \ \ \ \ \ \ \ \ \\[2pt]
s_j  = z_j. \ \ \ \ \ \ \ \ \ \ \ \\[2pt]
\end{array}
\right. $$
Then we have $0 \notin \{t_i - x_y, t_j - x_z, (t_i - x_y) \pm (t_j - x_z), s_i - y_i, s_j - y_j, (s_i - y_i) \pm (s_j - y_j),
t_i - x_{z_i}, t_j - x_{z_j}, (t_i - x_{z_i}) \pm (t_j - x_{z_j})\}$.
\end{itemize}
\end{theorem}

\begin{theorem}
\label{condiSESC}   Suppose that ${\cal C}$  is a $\overline{2}$-{\rm SC}$(3, q(q+w), q)$ in the form {\rm (iii)} on  $Z_q$,
and $E= \{ ( y , x_y ) \ | \ y \in Z_q \} \ \bigcup \ \{ ( s_i, t_i) \ | \ 1 \leq i \leq w\}$.
Then ${\cal C}$  is a $2$-SMIPPC$(3, q(q+w), q)$
provided that the following hold:\\
{\rm (I)}   There do not exist distinct $ 1 \leq i_1, i_2, i_3 \leq w$ and $y \in Z_q$,  such that
$$
\left\{\begin{array}{rl}
y = s_{i_1}, \ \ \ \ \ \ \ \ \ \ \ \ \ \ \ \ \ \ \ \ \ \ \ \ \ \ \ \ \ \ \  \\[2pt]
x_y = t_{i_2},  \ \ \ \ \ \ \ \ \ \ \ \ \ \ \ \ \ \ \ \ \ \ \ \ \ \ \ \ \ \   \\[2pt]
x_y -  y = t_{i_3} - s_{i_3},  \ \ \ \ \ \ \ \ \ \ \ \ \ \ \ \ \ \ \  \\[2pt]
(s_{i_2} + t_{i_3} - x_y, t_{i_1} + t_{i_3} - x_y) \in E.
\end{array}
\right.
$$
{\rm (II)}   There do not exist distinct $ y_1, y_2, y_3 \in Z_q$ and $ 1 \leq  i \leq w$,  such that
$$
\left\{\begin{array}{rl}
s_{i} = y_1, \ \ \ \ \ \ \ \ \ \ \ \ \ \ \ \ \ \ \ \ \ \ \ \ \ \ \ \ \ \ \  \\[2pt]
t_{i} =  x_{y_2}, \ \ \ \ \ \ \ \ \ \ \ \ \ \ \ \ \ \ \ \ \ \ \ \ \ \ \ \ \ \   \\[2pt]
t_{i} - s_{i} = x_{y_3} -  y_3,   \ \ \ \ \ \ \ \ \ \ \ \ \ \ \ \ \ \ \  \\[2pt]
(y_{2} + x_{y_3} - t_i, x_{y_1} + x_{y_3} - t_i) \in E.
\end{array}
\right.
$$
\end{theorem}
\begin{IEEEproof}  It is not difficult to check that ${\cal C}_D$ and ${\cal C}_S$ are codes with minimum distance $2$.
Assume that ${\cal C}$ is not a $2$-SMIPPC.
According to Theorem \ref{forbiconfi3}, there exists
${\cal C}_0 = \{ {\bf c}_1, {\bf c}_2 \} =  \{(a_1, b_1, e_1)^{T}, (a_2, b_2, e_2)^{T}\}  \subseteq {\cal C}$,
where $a_1 \neq a_2$, $b_1 \neq b_2$, and $e_1 \neq e_2$,
such that ${\sf desc}({\cal C}_0)\bigcap {\cal C}$ is of the following type:\\
\begin{eqnarray*}
{\bf c}_1 \ \  \ {\bf c}_2 \ \ \ {\bf c}_3 \ \ \ {\bf c}_4\ \ \  {\bf c}_5\ \ \ \ \\
{\sf desc}({\cal C}_0)\bigcap {\cal C}=
\left(
  \begin{array}{cc|cccccc}
    a_1 & a_2 & a_1 & a_1 & a_2\\
    b_1 & b_2 & b_1 & b_2 & b_1\\
    e_1 & e_2 & e_2 & e_1 & e_1\\
  \end{array}
\right)
\end{eqnarray*}
For convenience, suppose that ${\bf c}_3 = (a_1, b_1, e_2)^{T}$, ${\bf c}_4  = (a_1, b_2, e_1)^{T}$, ${\bf c}_5  = (a_2, b_1, e_1)^{T}$.

(1)   If ${\bf c}_1 \in {\cal C}_D$, then ${\bf c}_1 = (k, k + y, k + x_y)^{T}$, where $k, y \in Z_{q}$,
and ${\bf c}_3 = (k, k + y, e_2)^{T}$, ${\bf c}_4 = (k, b_2, k + x_y)^{T}$, ${\bf c}_5 = (a_2, k + y, k + x_y)^{T}$.

It is easy to see that $a_2 \neq k$.
Since ${\cal C}_D$ has minimum distance $2$, we have ${\bf c}_3, {\bf c}_4, {\bf c}_5 \in {\cal C}_S$.
Then there exist $1 \leq i_1, i_2, i_3 \leq w$ such that
${\bf c}_3 = (k, k + s_{i_1}, k + t_{i_1})^{T}$, ${\bf c}_4 = (k, k + s_{i_2}, k + t_{i_2})^{T}$, ${\bf c}_5 = (a_2, a_2 + s_{i_3}, a_2 + t_{i_3})^{T}$. Since ${\cal C}_S$ has minimum distance  $2$ and ${\bf c}_3, {\bf c}_4, {\bf c}_5 \in {\cal C}_S$, we have
$$
\left\{\begin{array}{c}
s_{i_1} \neq s_{i_2}, \ \ \ \ \ \ \ \ \ \ \ \ \ \ \ \  \ \ \ \ \ \ \ \ \ \ \ \ \ \ \  \ \ \ \ \ \ \ \
 \ \ \ \ \ \ \ \ \ \ \ \ \ \ \ \ \ \ \ \ \ \\[2pt]
t_{i_1} \neq t_{i_2},  \ \ \ \ \ \ \ \ \ \ \ \ \ \ \ \  \ \ \ \ \ \ \ \ \ \ \ \ \ \ \  \ \ \ \ \ \ \ \
 \ \ \ \ \ \ \ \ \ \ \ \ \ \ \ \ \ \ \ \ \  \\[2pt]
 k + t_{i_1} \neq a_2 + t_{i_3}  (  k + s_{i_1} = k + y = a_2 + s_{i_3}), \ \\[2pt]
k + s_{i_2} \neq a_2 + s_{i_3} ( k + t_{i_2} = k + x_y = a_2 + t_{i_3}).  \\[2pt]
\end{array}
\right.
$$
Obviously, $i_1 \neq i_2$. We can also derive $i_1 \neq i_3$, otherwise, if $i_1 = i_3$, then $k = a_2$, a contradiction.
Similarly, $i_2 \neq i_3$. So $i_1, i_2$ and $i_3$ are all distinct, and we have
$$
\left\{\begin{array}{rl}
k + y = k + s_{i_1}, \ \  \\[2pt]
e_2 =  k + t_{i_1},  \ \ \ \ \ \  \\[2pt]
b_2 =  k + s_{i_2},  \ \ \ \ \ \   \\[2pt]
k + x_y = k + t_{i_2},  \ \\[2pt]
k + y = a_2 + s_{i_3}, \ \\[2pt]
k + x_y = a_2 + t_{i_3}.  \\[2pt]
\end{array}
\right.
\Rightarrow
\left\{\begin{array}{rl}
 y = s_{i_1}, \ \ \ \ \ \ \ \ \ \ \ \ \ \\[2pt]
 x_y = t_{i_2},  \ \ \ \ \ \ \ \ \ \ \ \  \\[2pt]
x_y -  y = t_{i_3} - s_{i_3},  \\[2pt]
a_2 =  k + x_y- t_{i_3},  \\[2pt]
b_2 =  k + s_{i_2},  \ \ \ \ \ \  \ \\[2pt]
e_2 =  k + t_{i_1}.  \ \ \ \ \ \ \ \\[2pt]
\end{array}
\right.
$$
Then ${\bf c}_2 = (a_2, b_2, e_2)^{T} = (k + x_y- t_{i_3}, k + s_{i_2}, k + t_{i_1})^{T}.$

(1.1)   If ${\bf c}_2 \in {\cal C}_D$, then there exists $z \in Z_q$ such that
${\bf c}_2 = (k + x_y - t_{i_3}, k + x_y - t_{i_3} + z, k + x_y - t_{i_3} + x_z)^{T}$. So we  have
$$\left\{\begin{array}{rl}
 k + s_{i_2} =  k + x_y - t_{i_3} + z, \ \\[2pt]
 k + t_{i_1} =  k + x_y - t_{i_3} + x_z.   \\[2pt]
\end{array}
\right.
\Rightarrow
\left\{\begin{array}{rl}
z=  s_{i_2} + t_{i_3} - x_y, \ \\[2pt]
x_z = t_{i_1} + t_{i_3} - x_y. \\[2pt]
\end{array}
\right. \ \ \ \
$$
a contradiction to condition (I). So this case is impossible.

(1.2)   If ${\bf c}_2 \in {\cal C}_S$, then there exists $1 \leq i_4 \leq w$ such that
${\bf c}_2 = (k + x_y - t_{i_3}, k + x_y - t_{i_3} + s_{i_4}, k + x_y - t_{i_3} + t_{i_4})^{T}$. So we  have
$$\left\{\begin{array}{rl}
 k + s_{i_2} =  k + x_y - t_{i_3} + s_{i_4},  \\[2pt]
 k + t_{i_1} =  k + x_y - t_{i_3} + t_{i_4}. \  \\[2pt]
\end{array}
\right.
\Rightarrow
\left\{\begin{array}{rl}
s_{i_4}=  s_{i_2} + t_{i_3} - x_y,    \\[2pt]
t_{i_4} = t_{i_1} + t_{i_3} - x_y.  \  \\[2pt]
\end{array}
\right. \ \ \ \
$$
a contradiction to condition (I). So this case is impossible.

(2)   If ${\bf c}_1 \in {\cal C}_S$, then ${\bf c}_1 = (k, k + s_i, k + t_i)^{T}$, where $1 \leq i \leq w$,
and ${\bf c}_3 = (k, k + s_i, e_2)^{T}$, ${\bf c}_4 = (k, b_2, k + t_i)^{T}$, ${\bf c}_5 = (a_2, k + s_i, k + t_i)^{T}$.

It is easy to see that $a_2 \neq k$.
Since ${\cal C}_S$ has minimum distance $2$, we have ${\bf c}_3, {\bf c}_4, {\bf c}_5 \in {\cal C}_D$.
Then there exist $y_1, y_2, y_3 \in Z_q$ such that ${\bf c}_3 = (k, k + y_1, k + x_{y_1})^{T}$, ${\bf c}_4 = (k, k + y_2, k + x_{y_2})^{T}$, ${\bf c}_5 = (a_2, a_2 + y_{3}, a_2 + x_{y_3})^{T}$. Since ${\cal C}_D$ has minimum distance $2$ and ${\bf c}_3, {\bf c}_4, {\bf c}_5 \in {\cal C}_D$, we have
$$
\left\{\begin{array}{cc}
y_{1} \neq y_{2},\ \ \ \ \ \ \ \ \ \ \ \ \ \ \ \ \ \ \ \ \ \ \ \ \ \ \ \ \ \ \ \ \ \ \ \ \ \ \ \  \ \ \ \ \ \ \ \  \\[2pt]
x_{y_1} \neq x_{y_2},\ \ \ \ \ \ \ \ \ \ \ \ \ \ \ \ \ \ \ \ \ \ \ \  \ \ \ \ \ \ \ \ \ \ \ \ \ \ \ \  \ \ \ \ \ \   \\[2pt]
k + x_{y_1} \neq a_2 + x_{y_3} \ ( k + y_{1} = k + s_i = a_2 + y_3), \\[2pt]
k + y_{2} \neq a_2 + y_{3} \ ( k + x_{y_2} = k + t_i = a_2 + x_{y_3}). \\[2pt]
\end{array}
\right.
$$
If $y_1 = y_3$, then $k = a_2$, a contradiction. So, $y_1 \neq y_3$
Similarly, $y_2 \neq y_3$. So $y_1, y_2$ and $y_3$ are all distinct, and we have
$$
\left\{\begin{array}{rl}
k + s_i = k + y_{1}, \ \  \\[2pt]
e_2 =  k + x_{y_1},  \ \ \ \ \ \  \\[2pt]
b_2 =  k + y_{2},  \ \ \ \ \ \ \  \\[2pt]
k + t_i = k + x_{y_2},  \ \\[2pt]
k + s_i = a_2 + y_{3}, \ \\[2pt]
k + t_i = a_2 + x_{y_3}.  \\[2pt]
\end{array}
\right.
\Rightarrow
\left\{\begin{array}{c}
s_i = y_{1},  \\[2pt]
t_i = x_{y_2},   \\[2pt]
t_i - s_i = x_{y_3} - y_3,  \\[2pt]
a_2 =  k + t_i - x_{y_3}, \\[2pt]
b_2 =  k + y_{2},   \\[2pt]
e_2 =  k + x_{y_1}.   \\[2pt]
\end{array}
\right. \ \ \ \
$$
Then ${\bf c}_2 = (a_2, b_2, e_2)^{T} = (k + t_i - x_{y_3}, k + y_{2},  k + x_{y_1})^{T}.$

(2.1)   If ${\bf c}_2 \in {\cal C}_D$, then there exists $y_4 \in Z_q$, such that
${\bf c}_2 = (k + t_i - x_{y_3}, k + t_i - x_{y_3} + y_4, k + t_i - x_{y_3} + x_{y_4})^{T}$. So we can have
$$\left\{\begin{array}{rl}
k + y_{2} =   k + t_i - x_{y_3} + y_4, \ \ \ \\[2pt]
k + x_{y_1} =  k + t_i - x_{y_3} + x_{y_4}.   \\[2pt]
\end{array}
\right.
\Rightarrow
\left\{\begin{array}{rl}
y_4 =  y_{2} + x_{y_3} -  t_i, \ \ \ \\[2pt]
x_{y_4} = x_{y_1} + x_{y_3} -  t_i. \\[2pt]
\end{array}
\right. \ \ \ \
$$
a contradiction to condition (II). So this case is impossible.

(2.2)   If ${\bf c}_2 \in {\cal C}_S$, then there exists $1 \leq j \leq w$, such that
${\bf c}_2 = (k + t_i - x_{y_3}, k + t_i - x_{y_3} + s_j, k + t_i - x_{y_3} + t_j)^{T}$. So we can have
$$\left\{\begin{array}{rl}
k + y_{2} =   k + t_i - x_{y_3} + s_j, \  \\[2pt]
k + x_{y_1} =  k + t_i - x_{y_3} + t_{j}.   \\[2pt]
\end{array}
\right.
\Rightarrow
\left\{\begin{array}{rl}
s_j =  y_{2} + x_{y_3} -  t_i, \  \\[2pt]
t_{j} = x_{y_1} + x_{y_3} -  t_i. \\[2pt]
\end{array}
\right. \ \ \ \
$$
a contradiction to condition (II). So this case is impossible.

Therefore, ${\cal C}$ is a $2$-SMIPPC$(3, q(q+w), q)$.
\end{IEEEproof}

\subsection{The case $q \equiv 1, 5 \pmod 6$}   %

We now consider the case $q \equiv 1, 5 \pmod 6$. To simplify our discussion, let $x_i = 2i$, $0 \leq i \leq q-1$,
$s_{j_1} \neq s_{j_2}$,  $1 \leq j_1 \neq j_2 \leq w$, $t_j = 3s_j$, $1 \leq j \leq w$, in $D$ in the form {\rm (i)}
and $S$ in the form {\rm (ii)}, respectively. Then we have two new matrices:
\begin{eqnarray*}
 D_1 =
\left(
  \begin{array}{cccc}
     0   &  0  &  \cdots  &     0   \\
     0   &  1  &  \cdots  &    q-1  \\
     0   &  2  &  \cdots  &   2(q-1) \\
  \end{array}
\right),\ \ \ \  {\rm (iv)}
\end{eqnarray*}
\begin{eqnarray*}
S_1 =
\left(
  \begin{array}{cccc}
     0      &    0   &   \cdots    &   0   \\
     s_1    &   s_2  &    \cdots   &  s_{w}  \\
    3s_1    &  3s_2   &  \cdots    & 3s_{w} \\
  \end{array}
\right). \ \ \ \  {\rm (v)}
\end{eqnarray*}

Let \begin{eqnarray*}
{\cal C}_{D_1} = \{ {\bf c} + g \ | \ {\bf c} \in D_1, g \in Z_{q} \},\ \ \ \ \ \ \ \ \ \ \ \ \ \ \ \ \ \ \ \ \ \ \ \ \ \ \\
{\cal C}_{S_1} = \{ {\bf c} + g \ | \ {\bf c} \in S_1, g \in Z_{q} \}, \
{\cal C}_1 = {\cal C}_{D_1} \bigcup {\cal C}_{S_1}.
\   {\rm (vi)}
\end{eqnarray*}

It is easy to check that $D_1$ is a $(q, 3, 1)$-CDM. Let $A_1= \{ s_1, s_2, \ldots, s_{w}\}$, $A_2= \{ 2b \ | \ b \in A_1\}$,
and $A_3= \{ -3b \ | \ b \in A_1\}$.
Then for any $(a', b', e')^{T} \in {\cal C}_{S_1}$,
we can have $b' - a' \in A_1$, $e' - b' \in A_2$, and $a' - e' \in A_3$.

\begin{theorem}
\label{ConstruSC}  Suppose $q \equiv 1, 5 \pmod 6$. Then ${\cal C}_1$ in the form  {\rm (vi)} is a
$\overline{2}$-{\rm SC}$(3, q(q+w), q)$ on  $Z_q$
provided that the following hold:\\
{\rm (I)}  $s_i \neq 0$  for any positive integer $1 \leq i \leq w$.\\
{\rm (II)}   $s_i + s_j \neq 0$  always holds for any positive integers $1 \leq i < j \leq w$.
\end{theorem}
\begin{IEEEproof}   We apply Theorem \ref{equi8}. It is not difficult to check that
$|\{s_1, s_2,  \ldots, s_w\}| = |\{3s_1, 3s_2, \ldots, 3s_w\}| = |\{2s_1, 2s_2, \ldots, 2s_w\}| = w$
from the fact $q \equiv 1, 5 \pmod 6$.
For any two columns $(0, s_i, 3s_i)^{T}$ and $(0, s_j, 3s_j)^{T}$ in $S_1$, $1 \leq i \neq j \leq w$,
suppose $(0, y, 2y)^{T}$, $(0, z, 2z)^{T}$, $(0, y_i, 2y_i)^{T}$, $(0, y_j, 2y_j)^{T}$,
$(0, z_i, 2z_i)^{T}$, $(0, z_j, 2z_j)^{T} \in D_1$, where $ y, z, y_i, y_j, z_i, z_j \in Z_{q}$, such that
$$
\left\{\begin{array}{rl}
2s_i = y,  \ \  \\[2pt]
2s_j = z,  \ \ \\[2pt]
3s_i  = 2y_i,    \\[2pt]
3s_j  = 2y_j,  \\[2pt]
s_i  = z_i, \ \ \\[2pt]
s_j  = z_j. \ \ \\[2pt]
\end{array}
\right.$$
Then
 $3s_i - 2y = 3s_i - 4s_i = -s_i \neq 0$,

 $3s_j - 2z = 3s_j - 4s_j = -s_j \neq 0$,

$(3s_i - 2y) \pm (3s_j - 2z) = -(s_i \pm s_j) \neq 0$,

$s_i - y_i = s_i - \frac{3}{2}s_i = - \frac{1}{2}s_i \neq 0 $,

 $s_j - y_j = s_j - \frac{3}{2}s_j = - \frac{1}{2}s_j \neq 0 $,

$(s_i - y_i) \pm (s_j - y_j) = - \frac{1}{2}(s_i \pm s_j) \neq 0$,

$3s_i - 2z_i = 3s_i - 2s_i = s_i \neq 0$,

 $3s_j - 2z_j = 3s_j - 2s_j = s_j \neq 0$,

$(3s_i - 2z_i) \pm (3s_j - 2z_j) = s_i \pm s_j \neq 0$.\\
Then the conclusion comes from Theorem \ref{equi8}.
\end{IEEEproof}

\begin{theorem}
\label{ConstruSESC}  Suppose that $q \equiv 1, 5 \pmod 6$. Then ${\cal C}_1$ in the form {\rm (vi)} is a
$2$-SMIPPC$(3, q(q+w), q)$ on $Z_q$ provided that the following hold:\\
{\rm (I)}   $s_i \neq 0 $  for any positive integer $1 \leq i \leq w$.\\
{\rm (II)}   $s_i + s_j \neq 0$  always holds for any positive integers $1 \leq i < j \leq w$.\\
{\rm (III)}  There does not exist an element $b \in Z_q$ such that
$b, \frac{2b}{3}, \frac{b}{2} \in A_1 = \{ s_1, s_2, \ldots, s_{w}\}$ and $13b = 0$.
\end{theorem}
\begin{IEEEproof}  According to Theorem \ref{ConstruSC}, we know that ${\cal C}_1$ is a $\overline{2}$-{\rm SC}.
Assume that ${\cal C}$ is not a $2$-SMIPPC, then one of conditions {\rm (I)} and {\rm (II)} of  Theorem \ref{condiSESC} does not hold.

$(1)$ Assume that condition {\rm (I)} of Theorem \ref{condiSESC} does not hold.
Then there  exist distinct $ 1 \leq i_1, i_2, i_3 \leq w$ and $y \in Z_q$ such that
\begin{eqnarray*}
&&\left\{\begin{array}{rl}
y = s_{i_1}, \ \ \ \ \ \ \ \ \ \ \ \ \ \ \ \ \ \ \ \ \ \ \ \ \ \ \ \ \ \ \ \ \ \ \ \ \  \\[2pt]
2y = 3s_{i_2},  \ \ \ \ \ \ \ \ \ \ \ \ \ \ \ \ \ \ \ \ \ \ \ \ \ \ \ \ \ \  \ \ \ \  \\[2pt]
y = 2s_{i_3}, \ \ \ \ \ \ \ \ \ \ \ \ \ \ \ \ \ \ \  \ \ \ \ \ \ \ \ \ \ \ \ \ \ \ \  \\[2pt]
(s_{i_2} + 3s_{i_3} - 2y, 3s_{i_1} + 3s_{i_3} - 2y) \in E.
\end{array}
\right.\\
\Rightarrow&&
\left\{\begin{array}{rl}
y = s_{i_1},   \ \ \ \ \ \  \\[2pt]
\frac{2}{3}y = s_{i_2},  \ \ \ \ \  \\[2pt]
\frac{1}{2}y = s_{i_3}, \ \ \ \ \  \\[2pt]
(\frac{1}{6}y, \frac{5}{2}y) \in E.
\end{array}
\right.
\end{eqnarray*}
where $E= \{ ( y' , 2y' ) \ | \ y' \in Z_q \} \ \bigcup \ \{ ( s', 3s') \ | \ s' \in A_1\}$.
This means that $y, \frac{2y}{3}, \frac{y}{2} \in A_1$, and $(\frac{1}{6}y, \frac{5}{2}y) \in E$.

$(1.1)$  If $(\frac{1}{6}y, \frac{5}{2}y) \in \{ ( y' , 2y' ) \ | \ y' \in Z_q \}$, then $\frac{2}{6}y = \frac{5}{2}y$,
which implies $13y = 0$, a contradiction to condition {\rm (III)}.

$(1.2)$  If $(\frac{1}{6}y, \frac{5}{2}y) \in \{ ( s', 3s') \ | \ s' \in A_1\}$, then $\frac{3}{6}y = \frac{5}{2}y$,
which implies $y = 0$, a contradiction to $0 \notin A_1$.

$(2)$ Assume that condition {\rm (II)} of Theorem \ref{condiSESC} does not hold.
Then there exist distinct $ y_1, y_2, y_3 \in Z_q$ and $ 1 \leq  i \leq w$  such that
\begin{eqnarray*}
&&\left\{\begin{array}{rl}
s_{i} = y_1, \ \ \ \ \ \ \ \ \ \ \ \ \ \ \ \ \ \ \ \ \ \ \ \ \ \ \ \ \ \ \ \ \ \ \ \   \\[2pt]
3s_{i} =  2y_2, \ \ \ \ \ \ \ \ \ \ \ \ \ \ \ \ \ \ \ \ \ \ \ \ \ \ \ \ \ \  \ \ \ \    \\[2pt]
2s_{i} = y_3,  \ \ \ \ \ \ \ \ \ \ \ \ \ \ \ \ \ \ \  \ \ \ \ \ \ \ \ \ \ \ \ \ \ \ \ \\[2pt]
(y_{2} + 2y_3 - 3s_i, 2y_1 + 2y_3 - 3s_i) \in E.
\end{array}
\right.\\
\Rightarrow &&
\left\{\begin{array}{rl}
y_1 = s_{i} , \ \ \ \ \ \  \ \ \\[2pt]
\frac{2}{3}y_2 =s_{i}, \ \ \ \ \ \    \\[2pt]
\frac{1}{2}y_3 = s_{i},   \ \ \ \ \  \ \\[2pt]
(\frac{5}{2}s_i, 3s_i) \in E.
\end{array}
\right.
\end{eqnarray*}

$(2.1)$  If $(\frac{5}{2}s_i, 3s_i) \in \{ ( y', 2y' ) \ | \ y' \in Z_q \}$, then $5s_i = 3s_i$,
which implies $s_i = 0$, a contradiction to condition {\rm (I)}.

$(2.2)$  If $(\frac{5}{2}s_i, 3s_i) \in \{ ( s', 3s') \ | \ s' \in A_1\}$, then $\frac{15}{2}s_i = 3s_i$,
which implies $s_i = 0$, a contradiction to condition {\rm (I)}.

The above $(1)$ and $(2)$ show that conditions {\rm (I)} and {\rm (II)} of Theorem \ref{condiSESC} always hold, which implies that ${\cal C}_1$
is a $2$-SMIPPC from Theorem \ref{condiSESC}.
\end{IEEEproof}

Now, we use Theorem \ref{ConstruSESC} to construct optimal $2$-SMIPPC$(3, M, q)$s for
$q \equiv 1, 5 \pmod 6$.

\begin{lemma}
\label{ConstruSESC11}
If $q \equiv 1, 5 \pmod 6$ and $q \not\equiv 0 \pmod {13}$,
then there exists a   $2$-SMIPPC$(3, q^2 + \frac{q(q-1)}{2}, q)$.
\end{lemma}
\begin{IEEEproof}  Let ${\cal C}_1$ be in the form {\rm (vi)} and $A_1 = \{1, 2, \ldots, \frac{q-1}{2}\}$.
The conclusion comes from Theorem \ref{ConstruSESC}.
\end{IEEEproof}

\begin{lemma}
\label{ConstruSESC12}  If $q \equiv 13, 65 \pmod {78}$,
then there exists a  $2$-SMIPPC$(3, q^2 + \frac{q(q-1)}{2}, q)$.
\end{lemma}
\begin{IEEEproof}  Let $q=13r$. Suppose that ${\cal C}_1$ is  in the form {\rm (vi)}
and $A_1 = \{1, \ldots, 4r-1, 4r+1, \ldots, \frac{q-1}{2}, 9r\}$.
We want to show that conditions (I), (II), (III)   in Theorem \ref{ConstruSESC} are satisfied.
Obviously, conditions (I) and (II) hold. Assume that there exists an element $b \in Z_q$ such that
$b, \frac{2b}{3}, \frac{b}{2} \in A_1$ and $13b = 0$.  Then $b$ should be a multiple of $r$ and thus we have $b \in \{r, 2r, 3r, 5r, 6r, 9r\}$. Then

\begin{center}
\begin{tabular}{|c||c|c|c|c|c|c|}
  \hline
       $b$       &   $r$   & $2r$ & $3r$ & $5r$ & $6r$ & $9r$ \\  \hline
  $\frac{2b}{3}$ &   $5r$  & $10r$ & $2r$ & $12r$ & $4r$ & $6r$ \\ \hline
  $\frac{b}{2}$  &   $7r$  & $r$   & $8r$ & $9r$ & $3r$ & $11r$ \\
  \hline
\end{tabular}
\end{center}
\begin{center}
{\bf Table  1}
\end{center}

From Table 1, we know that for any $b \in \{r, 2r, 3r, 5r$, $6r, 9r\}$, one of the elements
$\frac{2b}{3}$ and $\frac{b}{2}$ is not contained in $A_1$, a contradiction to $b, \frac{2b}{3}, \frac{b}{2} \in A_1$.
Hence, condition (III) is satisfied.

The conclusion then comes from Theorem \ref{ConstruSESC}.
\end{IEEEproof}

Combining Theorem \ref{upperbound1}, Lemmas \ref{ConstruSESC11} and  \ref{ConstruSESC12}, we have

\begin{theorem}
\label{ConstruSESC1}
There exists an optimal $2$-SMIPPC$(3, q^2 + \frac{q(q-1)}{2}, q)$ for any $q \equiv 1, 5 \pmod 6$.
\end{theorem}

\subsection{The case  $q \equiv 0, 2 \pmod 6$}   %

Next, we deal with the case $q \equiv 0, 2 \pmod 6$. Let $s = q-1$, then $s \equiv 1, 5 \pmod 6$.

In order to describe our constructions, we introduce a new element $\infty \notin Z_s$, and for any $a \in Z_s$, we define
$$ a + \infty = \infty + a = a \cdot \infty = \infty \cdot a = \infty.
$$

We now define a code
$$\indent\indent\indent\indent\indent\indent\indent\indent\indent\indent{\cal C}_2^{'}= {\cal C}_2 \bigcup {\cal C}_{T_2} \bigcup \{ (\infty, \infty, \infty)^{T} \} \indent\indent\indent\indent\indent\indent\indent\indent\indent\indent\indent\indent\indent
\indent\indent \indent\indent\indent\indent {\rm (vii)}$$
on $Z_s \bigcup \{ \infty\}$, where $s_1, s_2, \ldots, s_{w}, m \in Z_s$,
\begin{eqnarray*}
&&D_2 =
\left(
  \begin{array}{cccc}
     0   &  0  &  \cdots  &     0   \\
     0   &  1  &  \cdots  &    s-1  \\
     0   &  2  &  \cdots  &   2(s-1) \\
  \end{array}
\right),\\
&&S_2 =
\left(
  \begin{array}{cccc}
     0      &    0   &   \cdots    &   0   \\
     s_1    &   s_2  &    \cdots   &  s_{w}  \\
     3s_1   &  3s_2  &  \cdots     & 3s_{w} \\
  \end{array}
\right),  \ \ \ \ \ \ \ \ \ \ \ \ \ \ \ \ {\rm (viii)}\\
&&T_2=
\left(
  \begin{array}{cccc}
     \infty   &     m    &     0         \\
       0      &   \infty  &    m    \\
       m     &     0     &  \infty   \\
  \end{array}
\right),
\end{eqnarray*}
\noindent
${\cal C}_{D_2} = \{ {\bf c} + g \ | \ {\bf c} \in D_2, g \in Z_{s} \}$,
${\cal C}_{S_2} = \{ {\bf c} + g \ | \ {\bf c} \in S_2, g \in Z_{s} \}$,
${\cal C}_{T_2} = \{ {\bf c} + g \ | \ {\bf c} \in {T_2}, g \in Z_{s} \}$, and
${\cal C}_2 = {\cal C}_{D_2} \bigcup {\cal C}_{S_2}$.

\begin{theorem}
\label{ConstruSC1}  ${\cal C}_2^{'}$ in the form {\rm (vii)} is a $\overline{2}$-{\rm SC}$(3, s(s+w+3)+1, q)$
provided that the following hold:\\
{\rm (I)}  $s_i \neq 0 $  for any positive integer $1 \leq i \leq w$.\\
{\rm (II)}  $s_i + s_j \neq 0$  always holds for any positive integers $1 \leq i < j \leq w$.\\
{\rm (III)}  $m \notin \bigcup_{i=1}^{3} A_i$.
\end{theorem}
\begin{IEEEproof}  According to  Theorem \ref{ConstruSC},
we know that ${\cal C}_2 = {\cal C}_{D_2} \bigcup {\cal C}_{S_2}$ is
a $\overline{2}$-{\rm SC}$(3, s(s+w), s)$ defined on $Z_s$.
Hence, in ${\cal C}_2$, $|{\cal A}_{g_1}^{j} \bigcap {\cal A}_{g_2}^{j}| \leq 1$ holds for any
positive integer $1 \leq j \leq 3$ and any distinct $g_1, g_2 \in Z_s$ from Lemma \ref{SC}.
Now we define
\begin{small}
\begin{eqnarray*}
{\cal B}_{g}^{j}=
\left\{\begin{array}{rl}
{\cal A}_{g}^{j}\bigcup \{(\infty, g - m)^{T}, (g + m, \infty)^{T}\},   &  \mbox{if}  \ g \in Z_s, \ j=1,3, \\[2pt]
{\cal A}_{g}^{j}\bigcup \{(\infty, g + m)^{T}, (g - m, \infty)^{T}\},   &  \mbox{if}  \ g \in Z_s, \ j=2, \\[2pt]
\{(i, i + m)^{T} | i \in Z_s\} \bigcup \{ (\infty, \infty)^{T}\},   &  \mbox{if}  \ g = \infty, \ j=1, 3, \\[2pt]
\{(i + m, i)^{T} | i \in Z_s\} \bigcup \{ (\infty, \infty)^{T}\},   &  \mbox{if}  \ g = \infty, \ j=2. \\[2pt]
\end{array}
\right.
\end{eqnarray*} \end{small}
According to Lemma \ref{SC}, in order to prove that ${\cal C}_2^{'}$  is a $\overline{2}$-{\rm SC},
it suffices to show that $|{\cal B}_{g_1}^{j} \bigcap {\cal B}_{g_2}^{j}| \leq 1$ holds for any
positive integer $1 \leq j \leq 3$, and any distinct $g_1, g_2 \in Z_s \bigcup \{\infty\}$.

Since for any  distinct $g_1, g_2 \in Z_s$,
$\{(\infty, g_1 - m)^{T}, (g_1 + m, \infty)^{T}\} \bigcap \{(\infty, g_2 - m)^{T}, (g_2 + m, \infty)^{T}\} = \emptyset$,
and $\{(\infty, g_1 + m)^{T}, (g_1 - m, \infty)^{T}\} \bigcap \{(\infty, g_2 + m)^{T}, (g_2 - m, \infty)^{T}\} = \emptyset$,
we have ${\cal B}_{g_1}^{j} \bigcap {\cal B}_{g_2}^{j} = {\cal A}_{g_1}^{j} \bigcap {\cal A}_{g_2}^{j}$ for any integer $1 \leq j \leq 3$,
which implies $|{\cal B}_{g_1}^{j} \bigcap {\cal B}_{g_2}^{j}| \leq 1$.

Next, since $m \notin \bigcup_{i=1}^{3} A_i$,
we know that for any $g \in Z_s$,

\indent \ \ \ \ \ ${\cal B}_{g}^{1} \bigcap {\cal B}_{\infty}^{1} = \{(g + m, g + 2m)^{T}\},$

\indent \ \ \ \ \ ${\cal B}_{g}^{2} \bigcap {\cal B}_{\infty}^{2} = \{(g + \frac{m}{2}, g - \frac{m}{2})^{T}\},$

\indent \ \ \ \ \ ${\cal B}_{g}^{3} \bigcap {\cal B}_{\infty}^{3} = \{(g - 2m, g - m)^{T}\}.$

Then  $|{\cal B}_{g}^{j} \bigcap {\cal B}_{\infty}^{j}| = 1$  holds for any integer $1 \leq j \leq 3$.

This completes the proof.
\end{IEEEproof}

\begin{theorem}
\label{ConstruSESC111}  ${\cal C}_2^{'}$ in the form {\rm (vii)} is a $2$-SMIPPC$(3, s(s+w+3)+1, q)$
provided that the following hold:\\
{\rm (I)} $s_i \neq 0 $  for any positive integer $1 \leq i \leq w$.\\
{\rm (II)}   $s_i + s_j \neq 0$  always holds for any positive integers $1 \leq i < j \leq w$.\\
{\rm (III)}  There does not exist an element $b \in Z_s$ such that
$b, \frac{2b}{3}, \frac{b}{2} \in A_1$ and $13b = 0$. \\
{\rm (IV)}   $m \notin \bigcup_{i=1}^{3} A_i$, $-\frac{m}{2} \notin A_2$, $-2m \notin A_3$, $m \neq 0$.
\end{theorem}
\begin{IEEEproof}  It is clear that ${\cal C}_2^{'}$ is a $\overline{2}$-{\rm SC} from Theorem \ref{ConstruSC1}.\\
Assume that ${\cal C}_2^{'}$ is not a $2$-SMIPPC.
According to Theorem \ref{forbiconfi3}, there exists
${\cal C}_0 = \{ {\bf c}_1, {\bf c}_2 \} =  \{(a_1, b_1, e_1)^{T}, (a_2, b_2, e_2)^{T}\} \subseteq {\cal C}_2^{'}$,
where $a_1 \neq a_2$, $b_1 \neq b_2$, and $e_1 \neq e_2$,
such that ${\sf desc}({\cal C}_0)\bigcap {\cal C}_2^{'}$ is of the following type:\\
\begin{eqnarray*}
{\bf c}_1 \ \  \ {\bf c}_2 \ \ \ {\bf c}_3 \ \ \ {\bf c}_4 \ \ \ {\bf c}_5\ \ \ \ \\
{\sf desc}({\cal C}_0)\bigcap {\cal C}_2^{'}=
\left(
  \begin{array}{cc|cccccc}
    a_1 & a_2 & a_1 & a_1 & a_2\\
    b_1 & b_2 & b_1 & b_2 & b_1\\
    e_1 & e_2 & e_2 & e_1 & e_1\\
  \end{array}
\right)
\end{eqnarray*}
For convenience, suppose that ${\bf c}_3 = (a_1, b_1, e_2)^{T}$,
${\bf c}_4  = (a_1, b_2, e_1)^{T}$, ${\bf c}_5  = (a_2, b_1, e_1)^{T}$.

(1) If ${\bf c}_1 \in {\cal C}_{D_2}$, then ${\bf c}_1 = (k, k + b, k + 2b)^{T}$, where $k, b \in Z_{s}$.

(1.1)  If $b \notin \{ m, -\frac{m}{2}\}$,
then ${\bf c}_3, {\bf c}_4, {\bf c}_5 \in {\cal C}_2 = {\cal C}_{D_2} \bigcup {\cal C}_{S_2}$,
and also ${\bf c}_2 \in {\cal C}_2$. According to the proofs of Theorems  \ref{condiSESC} and  \ref{ConstruSESC}, this case is impossible.

(1.2)  If $b = m$, then ${\bf c}_4 = (k, b_2, k + 2m)^{T}$.
Since $s \equiv 1, 5 \pmod 6$ and $m \neq 0$, we have $-2m \neq m$, which implies
${\bf c}_4 \notin {\cal C}_{T_2}$.  Since $-2m \notin A_3$, we can derive that  ${\bf c}_4 \notin {\cal C}_{S_2}$.
Hence ${\bf c}_4 \in {\cal C}_{D_2}$, which, together with the fact that ${\cal C}_{D_2}$  has minimum distance $2$,
implies ${\bf c}_4 = {\bf c}_1$, a contradiction.
So this case is impossible.

(1.3)   If $b = -\frac{m}{2}$, then  ${\bf c}_5 = (a_2, k -\frac{m}{2}, k - m)^{T}$.
Since $s \equiv 1, 5 \pmod 6$ and $m \neq 0$, we  have $-\frac{m}{2} \neq m$, which implies
${\bf c}_5 \notin {\cal C}_{T_2}$.  Since $-\frac{m}{2} \notin A_2$, we can derive that  ${\bf c}_5 \notin {\cal C}_{S_2}$.
Hence ${\bf c}_5 \in {\cal C}_{D_2}$, which implies ${\bf c}_5 = {\bf c}_1$, a contradiction.
So this case is impossible.

(2)  If ${\bf c}_1 \in {\cal C}_{S_2}$, then ${\bf c}_1 = (k, k + b, k + 3b)^{T}$,
where $k \in Z_{s},  b \in A_1 \subseteq Z_{s}$. Since $m \notin \bigcup_{i=1}^{3} A_i$,
we know that  ${\bf c}_3, {\bf c}_4, {\bf c}_5 \notin {\cal C}_{T_2}$, which implies
${\bf c}_3, {\bf c}_4, {\bf c}_5 \in {\cal C}_2 = {\cal C}_{D_2} \bigcup {\cal C}_{S_2}$,
and also ${\bf c}_2 \in {\cal C}_2$. According to the  proofs of Theorems  \ref{condiSESC} and  \ref{ConstruSESC}, this case is impossible.

(3)  If ${\bf c}_1 \in {\cal C}_{T_2}$, without loss of generality, we may assume that ${\bf c}_1 = (\infty, b, b + m)^{T}$.
Then ${\bf c}_3 = {\bf c}_1$, a contradiction. So this case is impossible.

(4) If ${\bf c}_1 = (\infty, \infty, \infty)^{T}$,
then ${\bf c}_3 = {\bf c}_1$, a contradiction. So this case is impossible.

According to (1)-(4), we know that ${\cal C}_2^{'}$ is a $2$-SMIPPC$(3, s(s+w+3) + 1, q)$.
\end{IEEEproof}

\begin{lemma}
\label{ConstruSESC21}  If $q \equiv 0 \pmod{6} \geq  12$ and $q \not\equiv 1 \pmod{13}$,
then there exists a  $2$-SMIPPC$(3, q^2 + \frac{q(q-1)}{2}, q)$.
\end{lemma}
\begin{IEEEproof}  Let ${\cal C}_2^{'}$ be in the form {\rm (vii)},
$A_1 = \{1, 2, \ldots, \frac{s-1}{2}\}$, and $m = -2$.
Obviously, conditions (I) and (II) of Theorem \ref{ConstruSESC111} are satisfied.
Since $q \not\equiv 1 \pmod{13}$,  $s = q-1 \not\equiv 0 \pmod{13}$.
Then, except the element $0 \in Z_s$,
there is no element $b \in Z_s$, such that $13b = 0$, but $0 \notin A_1$.
This implies that condition (III) of Theorem \ref{ConstruSESC111} is satisfied.
Now consider condition (IV) of  Theorem \ref{ConstruSESC111}.
Remember that  $A_2 = \{2b \ | \ b \in A_1\}$, $A_3 = \{-3b \ | \ b \in A_1\}$.

(1) Obviously, $m \neq 0$, and $-\frac{m}{2} = 1 \notin A_2$.

(2) $-2m = 4 \notin A_3$.  Assume not. Then there exists $b \in A_1$, such that $4 = -3b$.
Then $b = -\frac{4}{3}$. Since $s= q-1 \equiv 5 \pmod{6}$, we write $s=6h + 5$ for some integer $h \geq 1$.
Then $b = 4h+2$ and $A_1 = \{1, 2, \ldots, 3h+2\}$, which implies $b \notin A_1$,
a contradiction.

(3)  Obviously, $m = -2 \notin A_1 \bigcup A_2$. It suffices to show that  $m = -2 \notin A_3$.
Assume not. Then there exists $b \in A_1$, such that $-2 = -3b$.
Then $b = \frac{2}{3} = 4h+4$, which implies $b \notin A_1$, a contradiction.

The conclusion then comes from Theorem \ref{ConstruSESC111}.
\end{IEEEproof}

\begin{lemma}
\label{ConstruSESC22}   If $q \equiv 66 \pmod{78}$,
then there exists a  $2$-SMIPPC$(3, q^2 + \frac{q(q-1)}{2}, q)$.
\end{lemma}
\begin{IEEEproof}  Let $s = q-1 = 13r$, ${\cal C}_2^{'}$ be in the form {\rm (vii)},
$A_1 = \{1, \ldots, 4r-1, 4r+1, \ldots, \frac{s-1}{2}, 9r\}$, and $m = -2$.
Obviously, conditions (I) and (II) of Theorem \ref{ConstruSESC111} are satisfied.
Since $q \equiv 66 \pmod{78}$, $s = q-1 \equiv 65 \pmod{78}$,
then we can know condition (III) of Theorem \ref{ConstruSESC111} is satisfied from the proof of Lemma \ref{ConstruSESC12}.
Now consider condition (IV) of  Theorem \ref{ConstruSESC111}.
Remember that  $A_2 = \{2b \ | \ b \in A_1\}$, $A_3 = \{-3b \ | \ b \in A_1\}$.

(1) Obviously, $m \neq 0$, and $-\frac{m}{2} = 1 \notin A_2$.

(2)  $-2m = 4 \notin A_3$.  Assume not. Then there exists $b \in A_1$, such that $4 = -3b$.
Write $s=78h+65$. Then $r =6h+5$ and $b = -\frac{4}{3} = 52h+42$.
Since $\frac{s-1}{2}=39h+32$, it should hold that $b = 9r$, that is, $2h+3=0$, which is impossible.


(3)  Since $s \geq 65$, we have $r \geq 5$.
Then $s-2 = 13r -2 > 9r$, which implies $m = -2 \notin A_1$. Also, $s-2 = 13r-2 \neq 5r$,
which implies $m = -2 \notin A_2$.
It suffices to show that  $m = -2 \notin A_3$.
Assume not. Then there exists $b \in A_1$, such that $-2 = -3b$.
Then $b = \frac{2}{3} =52h+44$. Since $\frac{s-1}{2}=39h+32$, it should hold that $b = 9r$, that is, $2h+1=0$, which is impossible.

The conclusion then comes from Theorem \ref{ConstruSESC111}.
\end{IEEEproof}

\begin{lemma}
\label{ConstruSESC23}   If $q \equiv 2 \pmod{6} \geq  44$ and $q \not\equiv 1 \pmod{13}$,
then there exists a  $2$-SMIPPC$(3, q^2 + \frac{q(q-1)}{2}, q)$.
\end{lemma}
\begin{IEEEproof}  Let ${\cal C}_2^{'}$ be in the form {\rm (vii)},
$A_1 = \{1, 2, \ldots, \frac{s-1}{2}\}$, and $m = -10$. Obviously, conditions (I) and (II) of Theorem \ref{ConstruSESC111} are satisfied.
Since $q \not\equiv 1 \pmod{13}$, $s = q-1 \not\equiv 0 \pmod{13}$.
Then, except the element $0 \in Z_{s}$,
there is no element $b \in Z_s$, such that $13b = 0$, but $0 \notin A_1$.
This implies that condition (III) of Theorem \ref{ConstruSESC111} is satisfied.
Now consider condition (IV) of  Theorem \ref{ConstruSESC111}.
Remember that  $A_2 = \{2b \ | \ b \in A_1\}$, $A_3 = \{-3b \ | \ b \in A_1\}$.

(1) Obviously, $m \neq 0$, and $-\frac{m}{2} = 5 \notin A_2$.

(2) $-2m = 20 \notin A_3$.  Assume not. Then there exists $b \in A_1$, such that $20 = -3b$.
Then $b = -\frac{20}{3}$.
Write $s = 6h+1$ for some integer $h \geq 7$. Then $b = 4h -6$ and $A_1 = \{1, 2, \ldots, 3h\}$,
which implies $b \notin A_1$, a contradiction.

(3)  Since $s \geq 43$, we have $s - 10 = 6h-9$, and  $\frac{s-1}{2} = 3h$, which implies  $m = -10 \notin A_1$.
It is also clear that $m = -10 \notin A_2$. We show that  $m = -10 \notin A_3$.
Assume not. Then there exists $b \in A_1$, such that $-10 = -3b$.
Then $b = \frac{10}{3} = 4h+4$, which implies $b \notin A_1$,
a contradiction.

So, the conclusion comes from
Theorem \ref{ConstruSESC111}.
\end{IEEEproof}

\begin{lemma}
\label{ConstruSESC24}   If $q \equiv 14 \pmod{78} \geq  92$,
then there exists a  $2$-SMIPPC$(3, q^2 + \frac{q(q-1)}{2}, q)$.
\end{lemma}
\begin{IEEEproof}  Let $s = q-1 = 13r$, ${\cal C}_2^{'}$ be in the form {\rm (vii)},
$A_1 = \{1, \ldots, 4r-1, 4r+1, \ldots, \frac{s-1}{2}, 9r\}$, and $m = -10$.
Obviously, conditions (I) and (II) of Theorem \ref{ConstruSESC111} are satisfied.
Since $q \equiv 14 \pmod{78}$, $s = q-1 \equiv 13 \pmod{78}$,
then we can know that condition (III) of Theorem \ref{ConstruSESC111} is satisfied from the proof of Lemma \ref{ConstruSESC12}.
Now we consider condition (IV) of  Theorem \ref{ConstruSESC111}.

(1)  Obviously, $m \neq 0$, and $-\frac{m}{2} = 5 \notin A_2$.

(2) $-2m = 20 \notin A_3$.  Assume not. Then there exists $b \in A_1$, such that $20 = -3b$. Write $s=78h+13$.
Then $r = 6h+1$, and $b = -\frac{20}{3} = 52h+2$.
Since $\frac{s-1}{2} = 39h+6$, it should hold that $b =9r$,
that is, $2h+7=0$, which is impossible.

(3) Since $s \geq 91$, we can have $r \geq 7$.
Then  $s-10 = 13r -10 > 9r$, which implies $m = -10 \notin A_1$.

$m = -10 \notin A_2$. Assume not. Then there exists $b \in A_1$, such that $-10 = 2b$, which implies $b = -5 = s-5$.
Since $s \geq 91$, we have $s-5 > \frac{s-1}{2}$, which implies $b =9r$, that is, $ -10 = 2\cdot9r = 18r =5r$.
Hence $r = -2 = s-2 \equiv 5 \pmod{6}$, a contradiction.

It suffices to show that  $m = -10 \notin A_3$.
Assume not. Then there exists $b \in A_1$, such that $- 10 = -3b$.
Then $b = \frac{10}{3} = 52h+12$.
Since $\frac{s-1}{2} = 39h+6$, it should hold that $b =9r$,
then $- 10 = -3\cdot 9r = -r$, $r \equiv 10 \pmod s \equiv 10 \pmod {13r}$,  and $r= 10 \not\equiv 1 \pmod{6}$, a contradiction.

So the conclusion comes from Theorem \ref{ConstruSESC111}.
\end{IEEEproof}

\begin{lemma} \ \
\label{ConstruSESC25} There exists a  $2$-SMIPPC$(3, q^2 + \frac{q(q-1)}{2}, q)$ for any $q \in \{20, 26, 32, 38\}$.
\end{lemma}
\begin{IEEEproof} Let
\begin{eqnarray*}
&&A^{(20)} = \{1, 2, 3, 4, 5, 7, 8, 10, 13\}, \ m^{(20)}= 9,  \\
&&A^{(26)} = \{1, 2, 3, 4, 5, 6, 7, 8, 9, 10, 11, 13\},\ m^{(26)}= 24,  \\
&&A^{(32)} = \{1, 2, 3, 4, 5, 6, 7, 8, 9, 10, \\
&& \ \ \ \ \ \ \ \ \ \ \ \ \ \ \ \ \ \ \ \ \ \ \ \ \ \ \ \ 11, 12,  13, 15, 17\},\  m^{(32)}= 21,  \\
&&A^{(38)} = \{1, 2, 3, 4, 5, 6, 7, 8, 9, 10, 11,12, 13, \\
&& \ \ \ \ \ \ \ \ \ \ \ \ \ \ \ \ \ \ \ \ \ \ \ \ \ \ \ \ \  14, 15, 16, 17,19\}\  m^{(38)}= 27.
\end{eqnarray*}
For any $q \in \{20, 26, 32, 38\}$, let $s= q-1$, ${\cal C}^{(q)}$ be in the form {\rm (vii)},
$A_1 =A^{(q)}$, and $m = m^{(q)}$. Then the conclusion comes from Theorem \ref{ConstruSESC111}.
\end{IEEEproof}

The following result comes from Lemmas \ref{ConstruSESC21}-\ref{ConstruSESC25} and Theorem \ref{upperbound1}.

\begin{theorem}
\label{ConstruSESC2}
Suppose that  $q \equiv 0, 2 \pmod 6$, and $q \notin \{2, 6, 8, 14\}$, then there exists an optimal
 $2$-SMIPPC$(3, q^2 + \frac{q(q-1)}{2}, q)$.
\end{theorem}

For each $q \in \{6, 8\}$, we want to find the set $A_1$ and the element $m$ satisfying the conditions (I)-(IV) of Theorem \ref{ConstruSESC111}.
Unfortunately, we fail to do this. However, we can construct a $2$-SMIPPC$(3, q^2 + \frac{q(q-1)}{2}, q)$ for each $q \in \{6, 8\}$
by making a detailed analysis of the proof of Theorem \ref{ConstruSESC111}.

\begin{lemma}
\label{ConstruSESC26}
There exists a  $2$-SMIPPC$(3, 51, 6)$.
\end{lemma}
\begin{IEEEproof}  Let $s= 5$, ${\cal C}_2^{'}$ be in the form {\rm (vii)},
$A_1 = \{1, 2\}$, and $m = 3$. Then $A_2 = A_3 = \{2, 4\}$.
It is not difficult to check that $m \neq 0$, $-\frac{m}{2} = 1 \notin A_2$, $m \notin \bigcup_{i=1}^{3} A_i$,
and conditions (I), (II), (III) of Theorem \ref{ConstruSESC111} are satisfied.
According to the proof of Theorem \ref{ConstruSESC111}, it suffices to prove the following assertion:

There does not exist ${\cal C}_0 = \{ {\bf c}_1, {\bf c}_2 \} =  \{( k, k + 3, k + 1)^{T}$, $ (a_2, b_2, e_2)^{T}\} \subseteq {\cal C}_2^{'}$,
where $k \in Z_5$, $a_2, b_2, e_2 \in Z_5 \bigcup \{\infty\}$, $a_2 \neq k$, $b_2 \neq k + 3$, and $e_2 \neq k + 1$,
such that ${\sf desc}({\cal C}_0)\bigcap {\cal C}_2^{'}$ is of the following type:\\
\begin{eqnarray*}
 {\bf c}_1 \ \ \ \   {\bf c}_2 \ \ \ \  \ \ \ {\bf c}_3 \ \ \ \  \ \ \ {\bf c}_4 \ \ \ \  \ \ {\bf c}_5\ \ \ \ \ \ \ \ \\
{\sf desc}({\cal C}_0)\bigcap {\cal C}_2^{'}=
\left(
  \begin{array}{cc|cccccc}
    k      & a_2 \ \ & k     & k     & a_2    \\
    k + 3  & b_2 \ \ & k + 3 & b_2   & k + 3  \\
    k + 1  & e_2 \ \ & e_2   & k + 1 & k + 1  \\
  \end{array}
\right),
\end{eqnarray*}
where ${\bf c}_3 = ( k, k + 3, e_2)^{T}$,
${\bf c}_4  = ( k, b_2, k + 1)^{T}$, ${\bf c}_5  = ( a_2, k + 3, k + 1)^{T}$.

Assume not. Obviously, ${\bf c}_3, {\bf c}_4, {\bf c}_5 \notin {\cal C}_{D_2}$, because of
the fact that ${\cal C}_{D_2}$  has minimum distance $2$ and ${\bf c}_1 \in {\cal C}_{D_2}$.
It is not difficult to see that ${\bf c}_3, {\bf c}_5 \notin {\cal C}_{S_2}$ since $3 \notin A_1 \bigcup A_2$, which implies
${\bf c}_3, {\bf c}_5 \in {\cal C}_{T_2}$. Hence ${\bf c}_3 = ( k, k + 3, \infty)^{T}$, ${\bf c}_5  = ( \infty, k + 3, k + 1)^{T}$,
and ${\bf c}_2 = ( \infty, \infty, \infty)^{T}$, which implies ${\bf c}_4  = ( k, \infty, k + 1)^{T}$.
Clearly, since $m=3 \neq -1$, we know that ${\bf c}_4  = ( k, \infty, k + 1)^{T} \notin {\cal C}_2^{'}$, a contradiction.

So, ${\cal C}_2^{'}$ is a $2$-SMIPPC$(3, 51, 6)$.
\end{IEEEproof}

\begin{lemma}
\label{ConstruSESC27}
There exists a  $2$-SMIPPC$(3, 92, 8)$.
\end{lemma}
\begin{IEEEproof}  Let $s= 7$, ${\cal C}_2^{'}$ be in the form {\rm (vii)},
$A_1 = \{1, 2, 4\}$, and $m = 3$. Then $A_2 = A_3 = \{1, 2, 4\}$.
It is not difficult to check that $m \neq 0$, $m \notin \bigcup_{i=1}^{3} A_i$ and
conditions (I), (II), (III) of Theorem \ref{ConstruSESC111} are satisfied.
According to the proof of Theorem \ref{ConstruSESC111}, it suffices to prove the following assertion:

There does not exist ${\cal C}_0 = \{ {\bf c}_1, {\bf c}_2 \} =  \{( k, k + b, k + 2b)^{T}, (a_2, b_2, e_2)^{T}\} \subseteq {\cal C}_2^{'}$,
where $k \in Z_7$, $b \in \{2, 3\}$, $a_2, b_2, e_2 \in Z_7 \bigcup \{\infty\}$, $a_2 \neq k$, $b_2 \neq k + b$, and $e_2 \neq k + 2b$,
such that ${\sf desc}({\cal C}_0)\bigcap {\cal C}_2^{'}$ is of the following type:\\
\begin{eqnarray*}
{\bf c}_1 \ \ \ \  \ {\bf c}_2 \ \ \ \  \ \ \ {\bf c}_3 \ \ \ \  \ \ \ {\bf c}_4 \ \ \ \  \ \ \ {\bf c}_5\ \ \ \ \ \ \ \ \ \\
 {\sf desc}({\cal C}_0)\bigcap {\cal C}_2^{'}=
\left(
  \begin{array}{cc|cccccc}
    k      & a_2 \ \ & k     & k     & a_2    \\
    k + b  & b_2 \ \ & k + b & b_2   & k + b  \\
    k + 2b  & e_2 \ \ & e_2   & k + 2b & k + 2b  \\
  \end{array}
\right),
\end{eqnarray*}
where ${\bf c}_3 = ( k, k + b, e_2)^{T}$,
${\bf c}_4  = ( k, b_2, k + 2b)^{T}$, ${\bf c}_5  = ( a_2, k + b, k + 2b)^{T}$.

Assume not. Since ${\cal C}_{D_2}$  has minimum distance $2$ and
 ${\bf c}_1 \in {\cal C}_{D_2}$, we know that ${\bf c}_3, {\bf c}_4, {\bf c}_5 \notin {\cal C}_{D_2}$.

$(1)$  The case $b = 2$. We can directly check that ${\bf c}_3, {\bf c}_5 \notin {\cal C}_{T_2}$ and
${\bf c}_4 \notin {\cal C}_{S_2}$, which implies ${\bf c}_3, {\bf c}_5 \in {\cal C}_{S_2}$ and
${\bf c}_4 \in {\cal C}_{T_2}$. Hence ${\bf c}_3 = ( k, k + 2, k + 6)^{T}$,
${\bf c}_5  = ( k + 1, k + 2, k + 4)^{T}$ and ${\bf c}_4  = ( k, \infty, k + 4)^{T}$,
which implies ${\bf c}_2  = ( k + 1, \infty, k + 6)^{T}$.
Obviously, since $m=3 \neq -5$, we know that ${\bf c}_2  = ( k + 1, \infty, k + 6)^{T} \notin {\cal C}_2^{'}$, a contradiction.

$(2)$  The case $b = 3$. It is not difficult to see that ${\bf c}_3, {\bf c}_5 \notin {\cal C}_{S_2}$, which implies
${\bf c}_3, {\bf c}_5 \in {\cal C}_{T_2}$. Hence ${\bf c}_3 = ( k, k + 3, \infty)^{T}$, ${\bf c}_5  = ( \infty, k + 3, k + 6)^{T}$,
and ${\bf c}_2 = ( \infty, \infty, \infty)^{T}$, which implies ${\bf c}_4  = ( k, \infty, k + 6)^{T}$.
Obviously, since $m=3 \neq 1$, we know that  ${\bf c}_4  = ( k, \infty, k + 6)^{T} \notin {\cal C}_2^{'}$, a contradiction.

So, ${\cal C}_2^{'}$ is a $2$-SMIPPC$(3, 92, 8)$.
\end{IEEEproof}

Now, we deal with the  case $q = 14$.

\begin{lemma}
\label{ConstruSESC28}
There exists a $2$-SMIPPC$(3, 287, 14)$.
\end{lemma}
\begin{IEEEproof}  We construct a (3, 287, 14) code ${\cal C}_3^{'}$ on $Z_{13} \bigcup \{ \infty\}$ as follows. Let
\begin{eqnarray*}
&&D_3 =\left(
  \begin{array}{cccc}
     0   &  0  &  \cdots  &     0   \\
     0   &  1  &  \cdots  &    12  \\
     0   &  2  &  \cdots  &   2 \times 12 \\
  \end{array}
\right),\\
&&S_3 =
\left(
  \begin{array}{cccccc}
     0      &    0   &   0    &   0   &   0    &   0   \\
     1    &   2  &    3   &  5 &   6    &    9  \\
     3   &  6  &  9     & 2 &   5    &   1   \\
  \end{array}
\right),\\
&&T_3=
\left(
  \begin{array}{cccc}
     \infty   &     6    &     0         \\
       0      &   \infty  &    4    \\
       7     &     0     &  \infty   \\
  \end{array}
\right),
\end{eqnarray*}
\begin{eqnarray*}
&&{\cal C}_{D_3} = \{ {\bf c} + g \ | \ {\bf c} \in D_3, g \in Z_{13} \},\\
&&{\cal C}_{S_3} = \{ {\bf c} + g \ | \ {\bf c} \in S_3, g \in Z_{13} \},\\
&&{\cal C}_{T_3} = \{ {\bf c} + g \ | \ {\bf c} \in T_3, g \in Z_{13} \},\\
&&{\cal C}_3 = {\cal C}_{D_3} \bigcup {\cal C}_{S_3},\\
&&{\cal C}_3^{'} = {\cal C}_3 \bigcup {\cal C}_{T_3} \bigcup \{ (\infty, \infty, \infty)^{T} \}.
\end{eqnarray*}
Let $A_1= \{ 1, 2, 3, 5, 6, 9\}$, $A_2= \{ 2b \ | \ b \in A_1\}$,
and $A_3= \{ -3b \ | \ b \in A_1\}$.

According to the proof of Lemma \ref{ConstruSESC12}, we know that ${\cal C}_3$  is
a $2$-SMIPPC defined on $Z_{13}$.  Similar to the proof of Theorem \ref{ConstruSC1}, we can prove that
${\cal C}_3^{'}$ is a  $\overline{2}$-{\rm SC} defined on $Z_{13} \bigcup \{ \infty\}$ (see Appendix \ref{Appendix4}). Now assume that ${\cal C}_3^{'}$
is not a $2$-SMIPPC. According to Theorem \ref{forbiconfi3}, there exists
${\cal C}_0 = \{ {\bf c}_1, {\bf c}_2 \} =  \{(a_1, b_1, e_1)^{T}, (a_2, b_2, e_2)^{T}\} \subseteq {\cal C}_3^{'}$,
where $a_1 \neq a_2$, $b_1 \neq b_2$, and $e_1 \neq e_2$,
such that ${\sf desc}({\cal C}_0)\bigcap {\cal C}_3^{'}$ is of the following type:\\
\begin{eqnarray*}
{\bf c}_1 \ \  \ {\bf c}_2 \ \ \ {\bf c}_3 \ \ \ {\bf c}_4 \ \ \ {\bf c}_5\ \ \ \ \   \\
 {\sf desc}({\cal C}_0)\bigcap {\cal C}_3^{'}=
\left(
  \begin{array}{cc|cccccc}
    a_1 & a_2 & a_1 & a_1 & a_2\\
    b_1 & b_2 & b_1 & b_2 & b_1\\
    e_1 & e_2 & e_2 & e_1 & e_1\\
  \end{array}
\right),
\end{eqnarray*}
where ${\bf c}_3 = (a_1, b_1, e_2)^{T}$,
${\bf c}_4  = (a_1, b_2, e_1)^{T}$, ${\bf c}_5  = (a_2, b_1$, $e_1)^{T}$.

(1) If ${\bf c}_1 \in {\cal C}_{D_3}$, then ${\bf c}_1 = (k, k + b, k + 2b)^{T}$, where $k, b \in Z_{13}$.
Since ${\cal C}_{D_3}$  has minimum distance $2$, we  have ${\bf c}_3, {\bf c}_4, {\bf c}_5 \notin {\cal C}_{D_3}$.

(1.1) If $b \notin \{ 4, 7, 10\}$, then ${\bf c}_3, {\bf c}_4, {\bf c}_5 \in {\cal C}_3$,
and also ${\bf c}_2 \in {\cal C}_3$, which contradict to the fact that ${\cal C}_3$ is a $2$-SMIPPC.
So this case is impossible.

(1.2) If $b = 4$, noting that $-2b = 5 \notin A_3 \bigcup \{ 6 \}$, we have ${\bf c}_4 \notin {\cal C}_{S_3} \bigcup {\cal C}_{T_3}$,
which implies  ${\bf c}_4 \notin {\cal C}_3^{'}$, a contradiction. So this case is impossible.

(1.3) If $b = 7$ or $10$, noting that $b \notin A_1 \bigcup \{ 4 \}$, we have ${\bf c}_3 \notin {\cal C}_{S_3} \bigcup {\cal C}_{T_3}$,
which implies  ${\bf c}_3 \notin {\cal C}_3^{'}$, a contradiction. So this case is impossible.

(2)  If ${\bf c}_1 \in {\cal C}_{S_3}$, then ${\bf c}_1 = (k, k + b, k + 3b)^{T}$,
where $k \in Z_{13},  b \in \{1, 2, 3, 5, 6, 9\}$. We can check that ${\bf c}_3, {\bf c}_4, {\bf c}_5 \notin {\cal C}_{T_3}$,
which implies ${\bf c}_3, {\bf c}_4, {\bf c}_5 \in {\cal C}_3$ and also ${\bf c}_2 \in {\cal C}_3$.
This is a contradiction to the fact that ${\cal C}_3$ is a $2$-SMIPPC. So this case is impossible.

(3)  ${\bf c}_1 \in {\cal C}_{T_3}$.  If ${\bf c}_1 = (\infty, k, k + 7)^{T}$ (or ${\bf c}_1 = (k, \infty, k - 6)^{T}$), $k \in Z_{13}$,
then ${\bf c}_3 = {\bf c}_1$, a contradiction. Similarly, if ${\bf c}_1 = ( k, k + 4, \infty)^{T}$, $k \in Z_{13}$, then ${\bf c}_4 = {\bf c}_1$,
a contradiction. So this case is impossible.

(4) If ${\bf c}_1 = (\infty, \infty, \infty)^{T}$,
then ${\bf c}_3 = {\bf c}_1$, a contradiction. So this case is impossible.

According to (1)-(4), we know that  ${\bf c}_1 \notin {\cal C}_3^{'}$, a contradiction.

So, ${\cal C}_3^{'}$ is a $2$-SMIPPC$(3, 287, 14)$.
\end{IEEEproof}

According to Theorem \ref{ConstruSESC2}, Lemmas \ref{ConstruSESC26}-\ref{ConstruSESC28}, and Theorem \ref{upperbound1},
we can derive the following result.

\begin{theorem}
\label{ConstruSESC2111}
There exists an optimal $2$-SMIPPC$(3, q^2 + \frac{q(q-1)}{2}, q)$ for any positive integer $q \equiv 0, 2 \pmod 6 > 2$.
\end{theorem}

Finally, we can also construct an optimal binary $2$-SMIPPC of length $3$.

\begin{lemma}
\label{ConstruSESC29}
There exists an optimal  $2$-SMIPPC$(3, 4, 2)$.
\end{lemma}
\begin{IEEEproof}  A  $2$-SMIPPC$(3, 4, 2)$ is listed below:
$${\cal C}=
\left(
  \begin{array}{cccc}
    0  & 1  & 0  & 0      \\
    0  & 0  & 1  & 0     \\
    0  & 0  & 0  & 1   \\
  \end{array}
\right)
$$

In order to show that the code $\cal C$ above is optimal, we only need  to prove that there is no
$2$-SMIPPC$(3, M, 2)$ with $M \geq 5$. Assume not. Suppose ${\cal C}^{'}$ is a
$2$-SMIPPC$(3, M, 2)$ with $M \geq 5$. Noting that $q=2$, we know that $M\leq8$.
Choose arbitrary $5$ codewords
${\bf c}_i = (a_i, b_i, e_i) \in \cal C^{'}$, $1 \leq i \leq 5$.  Then there must be two codewords
${\bf c}_i$ and ${\bf c}_j$, $1 \leq i \neq j \leq 5$, such that $d({\bf c}_i, {\bf c}_j) = 3$.
We may assume that $d({\bf c}_1, {\bf c}_2) = 3$, $a_1 = 0$ and $a_2 = 1$.
Hence desc$( \{ {\bf c}_1, {\bf c}_2\}) = \{0, 1\} \times \{0, 1\} \times \{0, 1\}$.

Now, we are going to show that desc$( \{ {\bf c}_3, {\bf c}_4, {\bf c}_5\})$ $= \{0, 1\} \times \{0, 1\} \times \{0, 1\}$.
If $a_3 = a_4 = a_5 = 0$, then $\{{\bf c}_1, {\bf c}_3, {\bf c}_4$, ${\bf c}_5\} = \{(0, 0, 0)^{T}, (0, 0, 1)^{T}$, $
(0, 1, 0)^{T}, (0, 1, 1)^{T}\}$. Hence desc$( \{ (0, 0, 0)^{T}, (0, 1, 1)^{T}\} )=$desc$( \{(0, 0,  1)^{T},  (0, 1, 0)^{T} \} )$,
while $\{ (0, 0, 0)^{T}, (0, 1, 1)^{T}\} \bigcap \{ (0, 0,  1)^{T}, (0, 1, 0)^{T}\} = \emptyset$,
a contradiction to the definition of a $2$-SMIPPC.
So, it is impossible that $a_3 = a_4 = a_5 = 0$. Similarly, it is impossible that $a_3 = a_4 = a_5 = 1$.
This means that $\{ a_3, a_4, a_5 \} = \{0, 1\}$. Similarly, we can prove that $\{ b_3, b_4, b_5 \} = \{0, 1\}$ and
$\{ e_3, e_4, e_5 \} = \{0, 1\}$.
So, desc$( \{ {\bf c}_3, {\bf c}_4, {\bf c}_5\}) = \{0, 1\} \times \{0, 1\} \times \{0, 1\}$, which implies
desc$( \{ {\bf c}_3, {\bf c}_4, {\bf c}_5\})$ = desc$( \{ {\bf c}_1, {\bf c}_2\} )$,
while $\{ {\bf c}_3, {\bf c}_4, {\bf c}_5\} \bigcap \{ {\bf c}_1, {\bf c}_2\} = \emptyset$,
a contradiction to the definition of a $2$-SMIPPC.

So, there does not exist a $2$-SMIPPC$(3, M, 2)$ with $M \geq 5$.
\end{IEEEproof}

\section{Conclusions}
\label{conclu}

In this paper, we first introduced a new notion of an SMIPPC for multimedia fingerprinting
to resist the averaging attack.
We then showed that the tracing algorithm based on a binary $t$-SMIPPC is more efficient than
that of a binary $t$-MIPPC. To be more precise, binary $t$-SMIPPCs can be used to  identify
at least one colluder with computational complexity linear to the product of the length of
the code and the number of authorized users. A composition construction for binary
$t$-SMIPPCs from $q$-ary $t$-SMIPPCs
was presented, which makes the research on $q$-ary $t$-SMIPPCs interesting.
We also obtained several infinite series of optimal $t$-SMIPPC$(2, M, q)$s with $t = 2, 3$.
It is worth mentioning that optimal
$2$-SMIPPC$(3, M, q)$s were constructed for each $q \equiv 0, 1, 2, 5 \pmod 6$.

\appendices

\section{}
\label{Appendix4}

\begin{theorem}
\label{1111111111}  ${\cal C}_3^{'}$  is a $\overline{2}$-{\rm SC}$(3, 287, 14)$
\end{theorem}
\begin{IEEEproof}  According to  Theorem \ref{ConstruSC},
we know that ${\cal C}_3 = {\cal C}_{D_3} \bigcup {\cal C}_{S_3}$ is
a $\overline{2}$-{\rm SC}$(3, 247, 13)$ defined on $Z_{13}$.
Hence, $|{\cal A}_{g_1}^{j} \bigcap {\cal A}_{g_2}^{j}| \leq 1$ holds for any
positive integers $1 \leq j \leq 3$ and any distinct $g_1, g_2 \in Z_{13}$ from Lemma \ref{SC}.
Now we define
\begin{small}
\begin{eqnarray*}
{\cal B}_{g}^{j}=
\left\{\begin{array}{rl}
{\cal A}_{g}^{j}\bigcup \{(\infty, g - 6)^{T}, (g + 4, \infty)^{T}\},  &  if\ g \in Z_{13},  j=1\\[2pt]
{\cal A}_{g}^{j}\bigcup \{(\infty, g + 7)^{T}, (g - 4, \infty)^{T}\}, &  if\ g \in Z_{13},  j=2\\[2pt]
{\cal A}_{g}^{j}\bigcup \{(\infty, g - 7)^{T}, (g + 6, \infty)^{T}\},  &  if \ g \in Z_{13},  j=3\\[2pt]
\{(i, i + 7)^{T} | i \in Z_{13}\} \bigcup \{ (\infty, \infty)^{T}\}, &  if\ g = \infty,  j=1\\[2pt]
\{(i + 6, i)^{T} | i \in Z_{13}\} \bigcup \{ (\infty, \infty)^{T}\}, &  if \ g = \infty,  j=2\\[2pt]
\{(i, i + 4)^{T} | i \in Z_{13}\} \bigcup \{ (\infty, \infty)^{T}\},  &  if \  g = \infty,  j=3\\[2pt]
\end{array}
\right.
\end{eqnarray*}
\end{small}
According to Lemma \ref{SC}, in order to prove that ${\cal C}_3^{'}$ is a $\overline{2}$-{\rm SC},
it suffices to show that $|{\cal B}_{g_1}^{j} \bigcap {\cal B}_{g_2}^{j}| \leq 1$ holds for any
positive integer $1 \leq j \leq 3$, and any distinct $g_1, g_2 \in Z_{13} \bigcup \{\infty\}$.

For any  distinct $g_1, g_2 \in Z_{13}$, we have
\begin{small}
\begin{eqnarray*}
\{(\infty, g_1 - 6)^{T}, (g_1 + 4, \infty)^{T}\} \bigcap \{(\infty, g_2 - 6)^{T}, (g_2 + 4, \infty)^{T}\} = \emptyset,\\
\{(\infty, g_1 + 7)^{T}, (g_1 - 4, \infty)^{T}\} \bigcap \{(\infty, g_2 + 7)^{T}, (g_2 - 4, \infty)^{T}\} = \emptyset,\\
\{(\infty, g_1 - 7)^{T}, (g_1 + 6, \infty)^{T}\} \bigcap \{(\infty, g_2 - 7)^{T}, (g_2 + 6, \infty)^{T}\} = \emptyset.
\end{eqnarray*}
\end{small}
Then ${\cal B}_{g_1}^{j} \bigcap {\cal B}_{g_2}^{j} = {\cal A}_{g_1}^{j} \bigcap {\cal A}_{g_2}^{j}$ for any integer $1 \leq j \leq 3$,
which implies $|{\cal B}_{g_1}^{j} \bigcap {\cal B}_{g_2}^{j}| \leq 1$. For any $g \in Z_{13}$,
we can also have
\begin{center}
${\cal B}_{g}^{1} \bigcap {\cal B}_{\infty}^{1} = \{(g + 7, g + 1)^{T}\},$\\[0.2cm]
${\cal B}_{g}^{2} \bigcap {\cal B}_{\infty}^{2} = \{(g + 3, g - 3)^{T}\},$\ \ \\[0.2cm]
${\cal B}_{g}^{3} \bigcap {\cal B}_{\infty}^{3} = \{(g - 8, g - 4)^{T}\}.$\\
\end{center}
Then   $|{\cal B}_{g}^{j} \bigcap {\cal B}_{\infty}^{j}| = 1$ for any integer $1 \leq j \leq 3$.

This completes the proof.
\end{IEEEproof}

\end{document}